\documentclass[journal,draft,onecolumn,11pt]{IEEEtran}
\usepackage{latexsym,,amssymb,amsmath,graphicx,epsf,cite,bbm,float}
\def\ninept{\def\baselinestretch{1.5}}
\ninept

\newcommand{\be}{\begin{equation}}
\newcommand{\ee}{\end{equation}}
\newcommand{\bea}{\begin{eqnarray}}
\newcommand{\eea}{\end{eqnarray}}
\newcommand{\nn}{\nonumber}

\newcommand{\vol}{ V}

\newcommand{\bts}{\mbox{$ \{b(t) \}$}}
\newcommand{\cts}{\mbox{$ \{c(t) \}$}}

\newcommand{\xts}{\mbox{$ \{x(t) \}$}}
\newcommand{\qts}{\mbox{$ \{\vec{q}(t) \}$}}
\newcommand{\yts}{\mbox{$ \{y(t) \}$}}
\newcommand{\nts}{\mbox{$ \{n(t) \}$}}

\newcommand{\llr}{\mathrm{LLR}}

\renewcommand{\vec}[1]{\mbox{\boldmath${#1}$}}

\newcommand{\ei}{\end{itemize}}
\newcommand{\bi}{\begin{itemize}}

\newcommand{\vh}{\vec{h}}
\newcommand{\Pa}{{\Gamma}}

\newcommand{\vl}{\vec{l}}

\newcommand{\ctr}{\hat{x}_{\mathrm{ctw}}(t)}
\newcommand{\pr}{\hat{x}_{\Pa_i}(t)}

\newcommand{\hbn}{\hat{x}_{\rho}(t)}

\newcommand{\vw}{\vec{w}}

\newcommand{\vy}{\mbox{$\vec{y}$}}
\newcommand{\vmy}{\mbox{$\bar{\vec{y}}$}}
\newcommand{\vmx}{\mbox{$\bar{\vec{x}}$}}

\newcommand{\vx}{\mbox{$\vec{x}$}}
\newcommand{\vv}{\mbox{$\vec{v}$}}
\newcommand{\vq}{\mbox{$\vec{q}$}}
\newcommand{\calW}{{\cal W}}
\newcommand{\calF}{{\cal F}}
\newcommand{\vqq}{\tilde{\vec{q}}}
\newcommand{\vf }{\vec{f}}
\newcommand{\vn}{\mbox{$\vec{n}$}}
\newcommand{\mx}{\mbox{$\bar{x}$}}

\newcommand{\mH}{\vec{H}}
\newcommand{\mHu}{\vec{H}_{r}}
\newcommand{\mQ}{\vec{Q}}

\newcommand{\MB}{\left[\begin{array}}
\newcommand{\ME}{\end{array}\right]}
\newcommand{\defi}{\stackrel{\bigtriangleup}{=}}

\begin{document}
\title{Linear MMSE-Optimal Turbo Equalization Using Context Trees}
\author{Nargiz Kalantarova, Kyeongyeon Kim, Suleyman S. Kozat, {\em
    Senior Member}, IEEE, and Andrew C. Singer, {\em Fellow}, IEEE
  \thanks{Andrew C. Singer (acsinger@illinois.edu) is
    with the Electrical and Computer Engineering Department at
    University of Illinois at Urbana-Champaign. Suleyman S. Kozat and
    Nargiz Kalantarova (\{skozat,nkalantarova\}@ku.edu.tr) are with
    the Electrical Engineering Department at Koc University, Istanbul,
    tel: +902123381684, +902123381490. Kim Kyeongyeon (adrianakky@gmail.com) is with Samsung Electronics, Gyeonggi-do, Republic of Korea.}} \maketitle
\begin{abstract}
Formulations of the turbo equalization approach to iterative
equalization and decoding vary greatly when channel knowledge is
either partially or completely unknown.  Maximum aposteriori
probability (MAP) and minimum mean square error (MMSE) approaches
leverage channel knowledge to make explicit use of soft information
(priors over the transmitted data bits) in a manner that is distinctly
nonlinear, appearing either in a trellis formulation (MAP) or inside
an inverted matrix (MMSE).  To date, nearly all adaptive turbo
equalization methods either estimate the channel or use a direct
adaptation equalizer in which estimates of the transmitted data are
formed from an expressly linear function of the received data and soft
information, with this latter formulation being most common.  We study
a class of direct adaptation turbo equalizers that are both adaptive
and nonlinear functions of the soft information from the decoder.  We
introduce piecewise linear models based on context trees that can
adaptively approximate the nonlinear dependence of the equalizer on
the soft information such that it can choose both the partition
regions as well as the locally linear equalizer coefficients in each
region independently, with computational complexity that remains of
the order of a traditional direct adaptive linear equalizer.  This
approach is guaranteed to asymptotically achieve the performance of
the best piecewise linear equalizer and we quantify the MSE
performance of the resulting algorithm and the convergence of its MSE
to that of the linear minimum MSE estimator as the depth of the
context tree and the data length increase.
\end{abstract}
\begin{center}
\bfseries EDICS Category:  MLR-APPL,  MLR-SLER, ASP-APPL.
\end{center}
\section{Introduction}
Iterative equalization and decoding methods, or so-called turbo
equalization \cite{dou95, ha97,be96}, have become increasingly popular
methods for leveraging the power of forward error correction to
enhance the performance of digital communication systems in which
intersymbol interference or multiple access interference are present.
Given full channel knowledge, maximum aposteriori probability (MAP)
equalization and decoding give rise to an elegant manner in which the
equlization and decoding problems can be (approximately) jointly
resolved \cite{dou95}.  For large signal constellations or when the
channel has a large delay spread resulting in substantial intersymbol
interference, this approach becomes computationally prohibitive and
lower complexity linear equalization strategies are often employed
\cite{GlLaLa97,tuhler, Song04}.  Computational complexity issues are
also exacerbated by the use of multi-input/multi-output (MIMO)
transmission strategies.  It is important to note that MAP and MMSE
formulations of the equalization component in such iterative receivers
make explicit use of soft information from the decoder that is a
nonlinear function of both the channel response and the soft
information \cite{tuhler}.  In a MAP receiver, soft information is
used to weight branch metrics in the receiver trellis \cite{Pr95}. In
an MMSE receiver, this soft information is used in the (recursive)
computation of the filter coefficients and appears inside of a matrix
that is inverted \cite{tuhler}. 

In practice, most communication systems lack precise channel knowledge
and must make use of pilots or other means to estimate and track the
channel if the MAP or MMSE formulations of turbo equalization are to
be used \cite{Pr95,tuhler}.  Increasingly, however, receivers based
on direct adaptation methods are used for the equalization component,
due to their attractive computational complexity \cite{GlLaLa97,
  Laot05, drost08}.  Specifically, the channel response is neither
needed nor estimated for direct adaptation equalizers, since the
transmitted data symbols are directly estimated based on the signals
received. This is often accomplished with a linear or decision
feedback structure that has linear complexity in the channel memory,
as opposed to the quadratic complexity of the MMSE formulation, and is
invariant to the constellation size \cite{Laot05}. A MAP receiver not
only needs a channel estimate, but also has complexity that is
exponential in the channel memory, where the base of the exponent is
the transmit constellation size \cite{Pr95}.  For example, underwater
acoustic communications links often have a delay spread in excess of
several tens to hundreds of symbol periods, make use of 4 or 16 QAM
signal constellations, and have multiple transmitters and receive
hydrophones \cite{drost08, Singer09}. In our experience, for such
underwater acoustic channels, MAP-based turbo equalization is
infeasible and MMSE-based methods are impractical for all but the most
benign channel conditions \cite{drost08}.  As such, direct-adaptation
receivers that form an estimate of the transmitted symbols as a linear
function of the received data, past decided symbols, and soft
information from the decoder have emerged as the most pragmatic
solution.  Least-mean square (LMS)-based receivers are used in
practice to estimate and track the filter coefficients in these
soft-input/soft-output decision feedback equalizer structures, which
are often multi-channel receivers for both SIMO and MIMO transmissions
\cite{Laot05, GlLaLa97,Song04}.

While such linear complexity receivers have addressed the
computational complexity issues that make MAP and MMSE formulations
unattractive or infeasible, they have also unduly restricted the
potential benefit of incorporating soft information into the
equalizer.  Although such adaptive linear methods may converge to
their ``optimal'', i.e., Wiener solution, they usually deliver
inferior performance compared to a linear MMSE turbo
receiver\cite{bidan:thesis}, since the Wiener solution for this
stationarized problem, replaces the time-varying soft information by
its time average \cite{tuhler, tuhler_thesis}.  It is inherent in the
structure of such adaptive approaches that an implicit assumption is
made that the random process governing the received data and that of
the soft-information sequence are both mean ergodic so that ensemble
averages associated with the signal and soft information can be
estimated with time averages. The primary source of performance loss
of these adaptive algorithms is due to their implicit use of the log
likelihood ratio (LLR) information from the decoder as stationary soft
decision sequence \cite{bidan:thesis}, whereas a linear MMSE turbo
equalizer considers this LLR information as nonstationary {\it a
  priori} statistics over the transmitted symbols \cite{tuhler}.

\begin{figure}
    \centerline{\epsfxsize=9.5cm \epsfbox{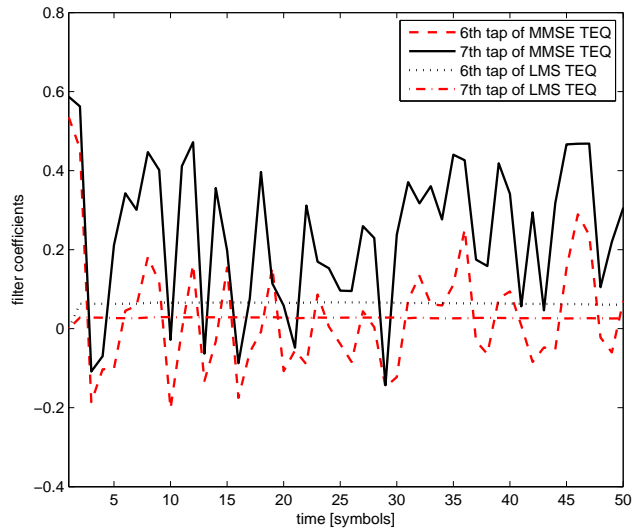}}
   \caption{An example of time varying filter coefficients of an MMSE turbo equalizer (TEQ) and steady state filter coefficients of an LMS turbo equalizer (TEQ) in a time invariant ISI channel $[0.227, 0.46, 0.688, 0.46, 0.227]$ at the second turbo iteration. ($SNR = 10dB$, feedforward filter length =$15$, step size = $0.001$, BPSK, random interleaver and $\frac{1}{2}$ rate convolutional code with constraint length of $3$ are used)}
      \label{fig:varyingfilter}
\end{figure}

Indeed, one of the strengths of the linear MMSE turbo equalizer lies
in its ability to employ a distinctly different linear equalizer for
each transmitted symbol \cite{tuhler,Song04}.  This arises from the
time-varying nature of the local soft information available to the
receiver from the decoder. Hence, even if the channel response were
known and fixed (i.e., time-invariant), the MMSE-optimal linear turbo
equlizer correponds to a set of linear filter coefficients that are
different for each and every transmitted symbol
\cite{tue02_5},\cite{tuhler}.  This is due to the presence of the soft
information inside a inverted matrix that is used to construct the
MMSE-optimal equalizer coefficients.  As a result, a time-invariant
channel will still give rise to a recursive formulation of the
equalizer coefficients that require quadratic complexity per output
symbol. As an example in Fig.~\ref{fig:varyingfilter}, we plot for a
time invariant channel the time varying filter coefficients of the
MMSE linear turbo equalizer, along with the filter coefficients of an
LMS-based, direct adaptation turbo equalizer that has converged to its
time invariant solution. This behavior is actually manifested due to
the nonlinear relationship between the soft information and the MMSE
filter coefficients.

In this paper, we explore a class of equalizers that maintain the
linear complexity adaptation of linear, direct adaptation equalizers
\cite{Laot05}, but attempt to circumvent the loss of this nonlinear
dependence of the MMSE optimal equalizer on the soft information from
the decoder \cite{tuhler}.  Specifically, we investigate an adaptive,
piecewise linear model based on context trees \cite{willems} that
partition the space of soft information from the decoder, such that
locally linear (in soft information space) models may be used. However
instead of using a fixed piecewise linear equalizer, the nonlinear
algorithm we introduce can adaptively choose the partitioning of the
space of soft information as well as the locally linear equalizer
coefficients in each region with computational complexity that remains
on the order of a traditional adaptive linear equalizer \cite{Pr95}.
The resulting algorithm can therefore successfully navigate the
short-data record regime, by placing more emphasis on lower-order
models, while achieving the ultimate precision of higher order models
as the data record grows to accommodate them. The introduced equalizer
can be shown to asymptotically (and uniformly) achieve the performance
of the best piecewise linear equalizer that could have been
constructed, given full knowledge of the channel and the received data
sequence in advance.  Furthermore, the mean square error (MSE) of this
equalizer is shown to convergence to that of the minimum MSE (MMSE)
estimator (which is a nonlinear function of the soft information) as
the depth of the context tree and data length increase.

Context trees and context tree weighting are extensively used in data
compression \cite{willems}, coding and data prediction
\cite{Ko08,HeSc97,TaMaVo01}. In the context of source coding and
universal probability assignment, the context tree weighting method is
mainly used to calculate a weighted mixture of probabilities generated
by the piecewise Markov models represented on the tree \cite{willems}.
In nonlinear prediction, context trees are used to represent piecewise
linear models by partitioning the space of past regressors
\cite{Ko08,TaMaVo01}, specifically for labeling the past
observations. Note that although we use the notion of context trees
for nonlinear modeling as in
\cite{willems,HeSc97,TaMaVo01,Ko07,koz_random_ctw}, our results and
usage of context trees differ from \cite{willems, HeSc97,TaMaVo01,
  Ko07} in a number of important ways. The ``context" used in our
context trees correspond to a spatial parsing of the soft information
space, rather than the temporal parsing as studied in \cite{willems,
  HeSc97,TaMaVo01}.  In addition, the context trees used here are
specifically used to represent the nonlinear dependency of equalizer
coefficients on the soft information. In this sense, as an example,
the time adaptation is mainly (in addition to learning) due to the
time variation of the soft information coming from the decoder, unlike
the time dependent learning in \cite{Ko07}. Hence, in here, we
explicitly calculate the MSE performance and quantify the difference
between the MSE of the context tree algorithm and the MSE of the
linear MMSE equalizer, which is the main objective.

The paper is organized as follows. In Section \ref{sec:system}, we
introduce the basic system description and provide the objective of
the paper. The nonlinear equalizers studied are introduced in Section
\ref{sec:alg}. In Section \ref{sec:alg}, we first introduce a
partitioned linear turbo equalization algorithm, where the
partitioning of the regions is fixed.  We continue in
Section~\ref{sec:alg_a} with the turbo equalization framework using
context trees, where the corresponding algorithm with the guaranteed
performance bounds is introduced. Furthermore, we provide the MSE
performance of all the algorithms introduced and compare them to the
MSE performance of the linear MMSE equalizer. The paper concludes with
numerical examples demonstrating the performance gains and the
learning mechanism of the algorithm.

\section{System Description \label{sec:system} }
Throughout the paper, all vectors are column vectors and represented
by boldface lowercase letters. Matrices are represented by boldface
uppercase letters. Given a vector $\vx$, $\|\vx\| = \sqrt{\vx^H \vx}$
is the $l_2$-norm, where $\vx^H$ is the conjugate transpose, $\vx^T$
is the ordinary transpose and $\vx^*$ is the complex conjugate.  For a
random variable $x$ (or a vector $\vx$), $E[x]=\bar{x}$ (or
$E[\vx]=\bar{\vx}$) is the expectation. For a vector $\vx$,
$\mathrm{diag}(\vx)$ is a diagonal matrix constructed from the entries
of $\vx$ and $x(i)$ is the $i$th entry of the vector. For a square
matrix $\vec{M}$, $\mathrm{tr}(\vec{M})$ is the trace. The sequences
are represented using curly brackets, e.g., $\xts$. $\bigcup_{i=1}^N
A_i$ denotes the union of the sets $A_i$, where $i=1,\hdots, N$. The
$\mathrm{vec}(.)$ operator stacks columns of a matrix of dimension
$m\times n$ into a $mn\times 1$ column vector \cite{Graham_matcal}.

\begin{figure}[t]
\centerline{\epsfxsize=12cm \epsfbox{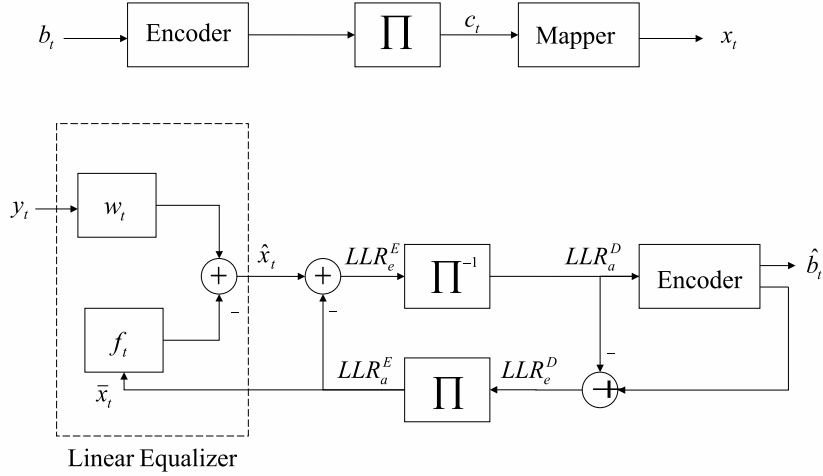}}
\caption{Block diagram for a bit interleaved coded modulation transmitter and receiver with a linear turbo equalizer.}
\label{fig:system}
\end{figure}

The block diagram of the system we consider with a linear turbo
equalizer is shown in Fig. \ref{fig:system}. The information bits
$\bts$ are first encoded using an error correcting code (ECC) and then
interleaved to generate $\cts$. The interleaved code bits $\cts$ are
transmitted after symbol mapping, e.g., $x(t) = (-1)^{c(t)}$ for BPSK
signaling, through a baseband discrete-time channel with a
finite-length impulse response $\{h(t)\}$, $t=0, 1, \hdots, M-1$,
represented by $\vh \defi [h(M-1),\ldots,h(0)]^T$. The communication
channel $\vh$ is unknown. The transmitted signal $\{x(t)\}$ is assumed
to be uncorrelated due to the interleaver. The received signal $y(t)$
is given by
\[
y(t) \defi \sum_{k=0}^{M-1} h(k) x(t-k)+n(t),
\]
where $\{n(t)\}$ is the additive complex white Gaussian noise with
zero mean and circular symmetric variance $\sigma_n^2$.  If a linear
equalizer is used to reduce the ISI, then the estimate of the desired
data $x(t)$ using the received data $y(t)$ is given by
\[
\hat{x}(t) = \vw^T(t) [\vec{y}(t)-\vmy(t)] + \mx(t),
\]
where $\vw(t) = [w(t,N_2),\ldots,w(t,-N_1)]^T$ is length $N=N_1+N_2+1$
linear equalizer, $\vy(t) \defi [ y(t-N_{2}),\ldots,y(t+N_{1})]^T$ and
note that we use negative indices with a slight abuse of
notation. The received data vector $\vy(t)$ is given by $\vy(t)
= \mH \vx(t) +\vn(t)$, where $\vx(t) \defi
[x(t-M-N_2+1),\ldots,x(t+N_{1})]^T$ and $\mH \in \mathbbm{C}^{N \times
  (N+M-1)}$
\[
\mH \defi \MB{cccccccc} h(M-1) & h(M-2) & \ldots & h(0) & 0 & \ldots  & & 0 \\
                    0 & h(M-1) & h(M-2) & \ldots & h(0) & 0 & \ldots &   0  \\
                    \ddots & \ddots & \ddots & \ddots & \ddots & \ddots &  \ddots &   \ddots \\
                       0 & \ldots & &  0 & h(M-1) & h(M-2) & \ldots & h(0)
                 \ME
\]
is the convolution matrix corresponding to $\vh =
[h(M-1),\ldots,h(0)]^T$, the estimate of $x(t)$ can be written as
\be
\hat{x}(t) = \vw^T(t) [\vy(t)-\mH \vmx(t)] +\mx(t), \label{eq:conv1}
\ee
given that the mean of the transmitted data is known.

However, in turbo equalization, instead of only using an equalizer,
the equalizer and decoder are jointly performed iteratively
at the receiver of Fig. \ref{fig:system}. The equalizer
computes the {\it a posteriori} information using the received
signal, transmitted signal estimate, channel convolution matrix (if
known) and {\it a priori} probability of the transmitted data. After
subtracting the {\it a priori} information, $\llr_a^E$,
and de-interleaving the extrinsic information $\llr_e^E$, a soft
input soft output (SISO) channel decoder computes the extrinsic
information $\llr_e^D$ on coded bits,
which are fed back to the linear equalizer as {\it a priori}
information $\llr_a^E$ after interleaving.

As the linear equalizer, if one uses the linear MMSE equalizer for
$\vw(t)$, the mean and the variance of $x(t)$ are required to
calculate $\vw(t)$ and $\hat{x}(t)$. These quantities are computed
using the {\it a priori} information from the decoder as $\mx(t)
\defi E[x(t) : \{\llr_a^E(t)\}]$\footnote{With a slight abuse of
notation, the expression $E[x(t) : y(t)]$ is interpreted here and in
the sequel as the expectation of $x(t)$ with respect to the prior
distribution $y(t)$.} and $q(t) \defi E[x^2(t) :
\{\llr_a^E(t)\}]-\mx^2(t)$. As an example, for BPSK signaling, the
mean and variance are given as $\bar{x}(t) = \tanh(\llr_a^E(t)/2)$
and $ q(t) = 1- |\bar{x}(t)|^2$. However, to remove dependency of
$\hat{x}(t)$ to $\llr_a^E(t)$ due to using $\mx(t)$ and $q(t)$ in
\eqref{eq:conv1}, one can set $\llr_a^E(t)=0$ while computing
$\hat{x}(t)$, yielding $\mx(t) = 0$ and $q(t)=1$ \cite{tuhler}.
Then, the linear MMSE equalizer is given by \be \vw(t) = [\vv^H
(\sigma_n^2 \vec{I}+ \mHu \mQ(t) \mHu^H + \vv \vv^H)^{-1}]^T,
\label{eq:lmmse} \ee where $\mQ(t)=
E[(\vx(t)-\bar{\vx}(t))(\vx(t)-\bar{\vx}(t))^H:\{LLR_a^E
(t)\}]$ is a diagonal matrix (due to
uncorrelateness assumption on $x(t)$) with diagonal entries $\mQ(t)=
\mathrm{diag}(\vq(t))$, $\vq(t) \defi
[q(t-M-N_2+1),\ldots,q(t-1),q(t+1),\ldots, q(t+N_1)]^T$, $\vv \in
\mathbbm{C}^{N}$ is the $(M+N_2)$th column of $\mH$, $\mHu$ is the
reduced form of $\mH$ where the $(M+N_2)$th column is removed.
The linear MMSE equalizer in \eqref{eq:conv1}
yields
\begin{align}
\hat{x}(t) & = \vw^T(t) [\vy(t)-\mH \vmx(t)] \nn \\
          & = \vw^T(t) \vy(t) - \vf^T(t) \vmx(t), \label{eq:fin}
\end{align}
where $\vf(t) \defi \mH^T \vw(t)$.  In this sense the linear MMSE
equalizer can be decomposed into a feedforward filter $\vw(t)$
processing $\vy(t)$ and a feedback filter $\vf(t)$ processing
$\vmx(t)$. \\

\noindent
{\bf Remark 1:} Both the linear MMSE feedforward and feedback filters are
highly nonlinear functions of $\vq$, i.e.,
\begin{align}
\vw &= \calW(\vq) \defi [\vv^H
(\sigma_n^2 \vec{I}+ \mHu \; \mathrm{diag}(\vq) \; \mHu^H + \vv
\vv^H)^{-1}]^T, \label{eq:non}\\
\vf &= \calF(\vq) \defi \mH^T \calW(\vq), \nn
\end{align}
where $\calW(\cdot), \; \calF(\cdot) : \mathbbm{C}^{N+M-2} \rightarrow
\mathbbm{C}^N$. We point out that time variation in \eqref{eq:lmmse}
is due to the time variation in the vector of variances $\vq$,
(assuming $\vh$ is time-invariant). \\

To learn the corresponding feedforward and
feedback filters that are highly nonlinear functions of $\vq$, we use
piecewise linear models based on vector quantization and context trees
in the next section. The space spanned by $\vq$ is partitioned into
disjoint regions and a separate linear model is trained for each
region to approximate functions $\calW(\cdot)$ and $\calF(\cdot)$
using piecewise linear models.

Note that if the channel is not known or estimated, one can directly
train the corresponding equalizers in \eqref{eq:fin} using adaptive
algorithms such as in \cite{GlLaLa97,bidan:thesis} without channel
estimation or piecewise constant partitioning as done in this paper.
In this case, one directly applies the adaptive algorithms to feedforward and
feedback filters using the received data $\yts$ and the mean vector
$\vmx(t)$ as feedback without considering the soft decisions as {\it a
  priori} probabilities. Assuming stationarity of $\vmx(t)$, such an
adaptive feedforward and feedback filters have Wiener solutions
\cite{bidan:thesis}
\begin{align}
\vw & = [\vv^H (\sigma_n^2 \vec{I}+ \mHu E[\vec{Q}(t)] \mHu^H + \vv \vv^H)^{-1}]^T, \label{eq:perf} \\
\vf & = \vec{H}^T\vw. \nonumber
\end{align}
Note that assuming stationarity of the log likelihood ratios
\cite{bidan:thesis}, $E[\vec{Q}(t)]$ is constant in time, i.e., no
time index for $\vw$, $\vf$ in \eqref{eq:perf}.  When PSK signaling is
used such that $E[|x(t)|^2 : \{\llr_a^E(t)\}]=1$, the filter
coefficient vector in \eqref{eq:perf} is equal to the coefficient
vector of the MMSE equalizer in \cite{tuhler} with the time averaged
soft information, i.e., time average instead of an ensemble
mean. Comparing \eqref{eq:perf} and \eqref{eq:lmmse}, we observe that
using the time averaged soft information does not degrade equalizer
performance in the no {\it a priori} information, i.e., $\vec{Q}(t) =
\vec{I}$ or perfect {\it a priori} information, i.e., $\vec{Q}(t) =
\vec{0}$, cases. In addition, the performance degradation in moderate
ISI channels is often small \cite{tuhler} when perfect channel
knowledge is used. However, the performance gap increases in ISI
channels that are more difficult to equalize, even in the high SNR
region \cite{bidan:thesis}, since the effect of the filter time
variation increases in the high SNR region.
\begin{figure} [t]
   \centerline{\epsfxsize=9cm \epsfbox{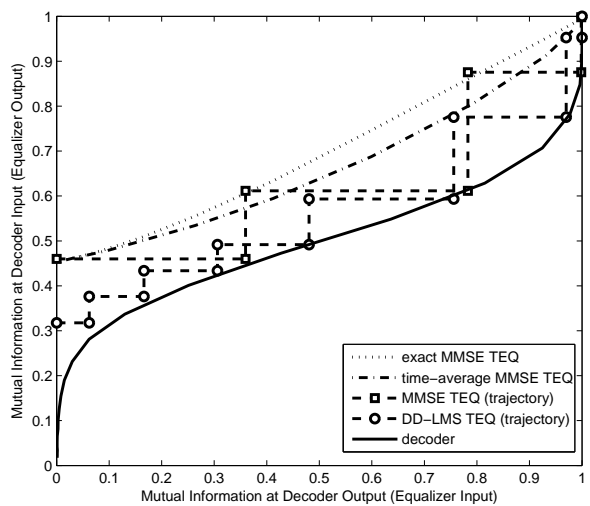}}
   \caption{The EXIT chart for the exact MMSE turbo equalizer, the LMS turbo equalizer and their trajectory in a time invariant ISI channel $[0.227, 0.46, 0.688, 0.46, 0.227]^T$. Here, we have $\mbox{SNR} = 10dB$, $N_1=9, N_2=5$, feedback filter length $N+M-1=19$, data block length $=8192$, training data length $=2048$, $\mu=0.001$, BPSK signaling, random interleaver and $\frac{1}{2}$ rate convolutional code with constraint length of $3$. }
   \label{fig:EXIT}
\end{figure}
Comparison of an exact MMSE turbo equalizer with/without channel estimation error and an MMSE turbo equalizer with the time averaged
soft variances (i.e. when an ideal filter for the converged adaptive turbo equalizer is used) via the EXIT
chart \cite{exit} is given in Fig. \ref{fig:EXIT}. As the adaptive turbo equalizer, a decision directed (DD) LMS turbo equalizer
is used in the data transmission period, while LMS is run on the received signals for the first turbo iteration
and on the received signals and training symbols for the rest of turbo iterations in the training period.
Note that the tentative decisions can be taken as the hard decisions
at the output of the linear equalizer or as the soft decisions from the total LLRs at the output of decoder.
When we consider nonideality, both of the MMSE turbo equalizer with channel estimation error and DD-LMS turbo equalizer loose
mutual information at the equalizer in first few turbo iterations
\footnote{This performance loss in the first few turbo iterations can
  fail the iterative process in low SNR region}.  Even though there is
a loss in mutual information at the equalizer in the first and second
turbo iteration due to using decision directed data or channel
estimation error, both algorithms follow their ideal performance at
the end.  (i.e., the DD LMS turbo equalizer can achieve the
performance of the time-average MMSE turbo equalizer as the decision
data gets more reliable). However, there is still a gap in achieved
mutual information between the exact MMSE turbo equalizer and the LMS
adaptive turbo equalizer except for the no {\em a priori} information
and perfect a priori information cases. Note that such a gap can make
an adaptive turbo equalizer become trapped at lower SNR region while
an MMSE turbo equalizer converges as turbo iteration increases.

To remedy this, in the next section, we introduce piecewise linear
equalizers to approximate $\calW(\cdot)$ and $\calF(\cdot)$. We first
discuss adaptive piecewise linear equalizers with a fixed partition of
$\mathbbm{C}^{N+M-2}$ (where $\vq \in \mathbbm{C}^{N+M-2}$). Then, we
 introduce adaptive piecewise linear equalizers using context trees
that can learn the best partition from a large class of possible
partitions of $\mathbbm{C}^{N+M-2}$.
\section{Nonlinear Turbo Equalization Using Piecewise Linear Models \label{sec:alg}}
\subsection{Piecewise Linear Turbo Equalization with Fixed Partitioning \label{sec:alg_fix}}
In this section, we divide the space spanned by $\vq
\in [0,1]^{N+M-2}$ (assuming BPSK signaling for notational simplicity)
into disjoint regions $\vol_k$, e.g., $[0,1]^{N+M-2} =\bigcup_{k=1}^K
\vol_{k}$ for some $K$ and train an independent linear equalizer in
each region $\vol_k$ to yield a final piecewise linear equalizer to
approximate $\vw = \calW(\vq)$ and $\vf=\calF(\vq)$. As an example,
given $K$ such regions, suppose a time varying linear equalizer is
assigned to each region as $\vw_k(t)$, $\vf_k(t)$, $k=1,\ldots,K$,
such that at each time $t$, if $\vq(t) \in \vol_k$, the estimate of
the received signal is given as
\begin{align}
& \hat{x}_k(t)  \defi \vw_k^T(t)\vy(t)-\vf_k^T(t)\vmx(t) \label{eq:1} \\
& \hat{x}(t)  = \hat{x}_k(t). \nn
\end{align}
We emphasize that the time variations in $\vw_k(t)$ and $\vf_k(t)$ in
\eqref{eq:1} are not due to the time variation in $\vq(t)$ unlike
\eqref{eq:fin}. The filters $\vw_k(t)$ and $\vf_k(t)$ are time varying
since they are produced by adaptive algorithms sequentially learning
the corresponding functions $\calW(\cdot)$ and $\calF(\cdot)$ in
region $\vol_k$. Note that if $K$ is large and the regions are dense
such that $\calW(\vq)$ (and $\calF(\vq)$) can be considered constant
in $\vol_k$, say equal to $\calW(\vq_k)$ for some $\vq_k$ in region
$\vol_k$, then if the adaptation method used in each region converges
successfully, this yields $\vw_k(t) \rightarrow \calW(\vq_k)$ and
$\vf_k(t) \rightarrow \calF(\vq_k)$ as $t \rightarrow \infty$. Hence,
if these regions are dense and there is enough data to learn the
corresponding models in each region, then this piecewise model can
approximate any smoothly varying $\calW(\vq)$ and $\calF(\vq)$
\cite{tree}.

In order to choose the corresponding regions
$\vol_{1},\ldots,\vol_{K}$, we apply a vector quantization (VQ)
algorithm to the sequence of $\qts$, such as the LBG VQ algorithm
\cite{book:gersho}. If a VQ algorithm with $K$ regions and
Euclidean distance is used for clustering, then the centroids and the
corresponding regions are defined as
\begin{align}
& \vqq_k \defi \frac{\sum_{t, \vq(t) \in \vol_k} \vq(t)}{\sum_{t, \vq(t) \in \vol_k}1}, \label{eq:vq1} \\
& \vol_k \defi \{ \vq:\|\vq-\vqq_k\| \leq \|\vq-\vqq_i\|, i = 1,\ldots,K,i \neq k \}, \label{eq:vq2}
\end{align}
where $\vq(t) = [q(t-M-N_2+1),\ldots,q(t-1),q(t+1),\ldots, q(t+N_1)]^T$, and $\vq \in
\mathbbm{C}^{N+M-2}$.
We emphasize that we use a VQ algorithm on $\qts$ to construct the
corresponding partitioned regions in order to concentrate on $\vq$
vectors that are in $\qts$ since $\calW(\cdot)$ and $\calF(\cdot)$
should only be learned around $\vq \in \qts$, not for all
$\mathbbm{C}^{N+M-2}$.
After the regions are constructed using the VQ
algorithm and the corresponding filters in each region are trained
with an appropriate adaptive method, the estimate of $x(t)$ at each time $t$ is given as
$\hat{x}(t) = \hat{x}_i(t)$ if $i = \arg\min_k \|\vq(t)-\vqq_k\|$.
\begin{figure}[t]
{\scriptsize
\begin{tabular}[t]{|l|}
\hline
\textbf{A Pseudo-code of Piecewise Linear Turbo Equalizer:}\\
\hline
{\bf \%  $1$st iteration:} \\
for $t=1,\ldots,n$:  \\
\hspace{0.3in}$\hat{x}(t) = \vw^{(1)T}(t)\vy(t)$, \\
\hspace{0.3in}if $t \leq T$: $e(t) = x(t)-\hat{x}(t)$, elseif $t > T$: $e(t) = Q(\hat{x}(t))-\hat{x}(t)$, $Q(\cdot)$ is a quantizer. \\
\hspace{0.3in}$\vw^{(1)}(t+1) = \vw^{(1)}(t)+ \mu e(t) \vy(t).$ \\
calculate $\vq(t)$ using the SISO decoder for $t >T$. \\
{\bf \% $2$nd iteration:} \\
apply LBG VQ algorithm to $\qts_{t>T}$ to generate $\vqq_k^{(2)}$,
$k=1,\ldots,K$. \\
for $k=1,\ldots,K$: $\vw_k^{(2)}(0) = \vw^{(1)}(n)$, where $\vw^{(1)}(n)$ is from 1st iteration. \hfill (line A) \\
for $t=1,\ldots,T$:  \\
\hspace{0.3in}for $k=1,\ldots,K$: \\
\hspace{0.6in}$e_k(t)=x(t)-\vw^{(2)T}_k(t)\vy(t)-\vf^{(2)T}_k(t)[\vec{I}-\mathrm{diag}(\vqq_k^{(2)})]^{1/2}\vx(t)$, \hfill (line B)\\
\hspace{0.6in}$\vw_k^{(2)}(t+1) = \vw_k^{(2)}(t)+ \mu e_k(t) \vy(t)$, $\vf_k^{(2)}(t+1) = \vf_k^{(2)}(t)+ \mu e_k(t)[\vec{I}-\mathrm{diag}(\vqq_k^{(2)})]^{1/2}\vx(t).$\\
for $t=T+1,\ldots,n$:  \\
\hspace{0.3in}$i = \arg \min_k \|\vq(t)-\vqq_k^{(2)}\|$, \\
\hspace{0.3in}$\hat{x}(t) = \vw^{(2)T}_i(t)\vy(t)-\vf^{(2)T}_i(t)\vmx(t)$, \\
\hspace{0.3in}$e(t) = Q(\hat{x}(t))-\hat{x}(t)$, \\
\hspace{0.3in}$\vw_i^{(2)}(t+1) = \vw_i^{(2)}(t)+ \mu e(t) \vy(t)$, $\vf_i^{(2)}(t+1) = \vf_i^{(2)}(t)+ \mu e(t) \vmx(t).$ \\
calculate $\qts_{t >T}$ using the SISO decoder. \\
{\bf \% $m$th iteration:} \\
apply LBG VQ algorithm to $\qts_{t>T}$ to generate $\vqq_k^{(m)}$,
$k=1,\ldots,K$. \\
for $k=1,\ldots,K$: $\vw_k^{(m)}(0)= \vw_j^{(m-1)}(n)$,  $\vf_k^{(m)}(0)= \vf_j^{(m-1)}(n)$, where $j = \arg\min_i \|\vqq_k^{(m)}-\vqq_i^{(m-1)}\|$. \hfill (line C) \\
for $t=1,\ldots,T$:  \\
\hspace{0.3in}for $k=1,\ldots,K$: \\
\hspace{0.6in}$e_k(t)=x(t)-\vw^{(m)T}_k(t)\vy(t)-\vf^{(m)T}_k(t)[\vec{I}-\mathrm{diag}(\vqq_k)]^{1/2}\vx(t)$, \\
\hspace{0.6in}$\vw_k^{(m)}(t+1) = \vw_k^{(m)}(t)+ \mu e_k(t) \vy(t)$, $\vf_k^{(m)}(t+1) = \vf_k^{(m)}(t)+ \mu e_k(t)[\vec{I}-\mathrm{diag}(\vqq_k^{(m)})]^{1/2}\vx(t).$\\
for $t=T+1,\ldots,n$,  \\
\hspace{0.3in}$i = \arg \min_k \|\vq(t)-\vqq_k^{(m)}\|$, \\
\hspace{0.3in}$\hat{x}(t) = \vw^{(m)T}_i(t)\vy(t)-\vf^{(m)T}_i(t)\vmx(t)$, \\
\hspace{0.3in}$e(t) = Q(\hat{x}(t))-\hat{x}(t)$, \\
\hspace{0.3in}$\vw_i^{(m)}(t+1) = \vw_i^{(m)}(t)+ \mu e(t) \vy(t)$, $\vf_i^{(m)}(t+1) = \vf_i^{(m)}(t)+ \mu e(t) \vmx(t).$ \\
calculate $\qts_{t >T}$ using the SISO decoder. \\
\hline
\end{tabular}}
\caption{A piecewise linear equalizer for turbo equalization. This algorithm requires $O(M+N)$ computations.}
\label{fig:piece}
\end{figure}

In Fig. \ref{fig:piece}, we introduce such a sequential piecewise
linear equalizer that uses the LMS update to train its equalizer
filters. Here, $\mu$ is the learning rate of the LMS updates. One can
use different adaptive methods instead of the LMS update, such as the
RLS or NLMS updates \cite{haykin96}, by only changing the filter
update steps in Fig. \ref{fig:piece}. The algorithm of
Fig. \ref{fig:piece} has access to training data of size $T$. After
the training data is used, the adaptive methods work in decision
directed mode \cite{haykin96}. Since there are no {\it a priori}
probabilities in the first turbo iteration, this algorithm uses an LMS
update to train a linear equalizer with only the feedforward filter,
i.e., $\hat{x}(t) = \vw^T(t) \vy(t)$, without any regions or
mean vectors.  Note that an adaptive feedforward linear filter
$\vw(t)$ trained on only $\yts$ without {\it a priori} probabilities
(as in the first iteration) converges to \cite{bidan:thesis}
(assuming zero variance in convergence)
\[
\lim_{t \rightarrow \infty} \vw(t)= \vw_o \defi  [\vv^H (\sigma_n^2 \vec{I}+ \mHu \mHu^H + \vv \vv^H)^{-1}]^T,
\]
which is the linear MMSE feedforward filter in \eqref{eq:lmmse} with
$\mQ(t) = \vec{I}$.

In the pseudo-code in Fig.~\ref{fig:piece}, the iteration numbers are
displayed as superscripts, e.g., $\vw^{(m)}_i(t)$, $\vf^{(m)}_i(t)$
are the feedforward and feedback filters for the $m$th iteration
corresponding to the $i$th region, respectively.  After the first
iteration when $\qts$ become available, we apply the VQ algorithm to
get the corresponding regions and the centroids. Then, for each
region, we run a separate LMS update to train a linear equalizer and
construct the estimated data as in \eqref{eq:1}. In the start of the
second iteration, in line A, each feedforward filter is initialized by
the feedforward filter trained in the first iteration. Furthermore,
although the linear equalizers should have the form
$\vw_k^T(t)\vy(t)-\vf_k^T(t)\vmx(t)$, since we have the correct
$\vx(t)$ in the training mode for $t=1,\ldots,T$, the algorithms are
trained using
$\vw_k^{T}(t)\vy(t)-\vf_k^{T}(t)[\vec{I}-\mathrm{diag}(\vqq_k)]^{1/2}\vx(t)$
in (line B), i.e., $\vx(t)$ is scaled using
$[\vec{I}-\mathrm{diag}(\vqq_k)]^{1/2}$, to incorporate the
uncertainty during training \cite{kim_singer}. After the second
iteration, in the start of each iteration, in line C, the linear
equalizers in each region, say $k$, are initialized using the filters
trained in the previous iteration that are closest to the $k$th
region, i.e., $j = \arg\min_i \|\vqq_k^{(m)}-\vqq_i^{(m-1)}\|$,
$\vw_k^{(m)}(0) = \vw_j^{(m-1)}(n)$, and $\vf_k^{(m)}(0) =
\vf_j^{(m-1)}(n)$.

Assuming large $K$ with dense regions, we have
$\vq(t) \approx \vqq_{k}$ when $\vq(t) \in \vol_k$. To get the vectors
that the LMS trained linear filters in region $k$ eventually converge, i.e.,
the linear MMSE estimators assuming stationary $\vmx$, we need to
calculate $E[(\vx(t)-\vmx(t))(\vx(t)-\vmx(t))^H : \vq(t)=\vqq_k ]$,
which is assumed to be diagonal due to the interleaving
\cite{bidan:thesis}, yielding
\begin{align*}
& E\{[\vx(t)-\vmx(t)][\vx(t)-\vmx(t)]^H : \vq(t)=\vqq_k \}  \\
& =E \bigg\{ E\big\{ [\vx(t)-\vmx(t)][\vx(t)-\vmx(t)]^H \: : \: \{\llr_a^E(t)\}, \vq(t)=\vqq_k \big\} \: : \:  \vq(t)=\vqq_k \bigg\} \\
& = \mathrm{diag} \big\{ [\vqq_k(1),\ldots,\vqq_k(M+N_2-1),1,\vqq_k(M+N_2),\ldots,\vqq_k(M+N-2)]\big\}
\end{align*}
due to the definition of $\vq(t)$ and assuming stationary distribution on
$\bar{x}(t)$.  This yields that the linear filters in region $k$
converge to
\begin{align}
& \lim_{t \rightarrow \infty} \vw_k(t) = \vw_{k,o} \defi [\vv^H (\sigma_n^2 \vec{I}+
\mHu \tilde{\mQ}_{k} \mHu^H + \vv \vv^H)^{-1}]^T,  \nn \\ &
\lim_{t \rightarrow \infty} \vf_k(t)=\vf_{k,o} \defi \mH^T\vw_{k,o},  \label{eq:applms}
\end{align}
where $\tilde{\mQ}_k \defi \mathrm{diag}(\vqq_{k})$, assuming zero
variance at convergence. Hence, at each time $t$, assuming
convergence, the difference between the MSE of the equalizer in
\eqref{eq:applms} and the MSE of the linear MMSE equalizer in
\eqref{eq:lmmse} is given by \be \| \vw_{k,o}^T \mHu \mQ(t) \mHu^H
\vw_{k,o}^*+ \sigma_n^2 \vw_{k,o}^T \vw_{k,o}^* - [1 - \vv^H
  (\sigma_n^2 \vec{I}+ \mHu \mQ(t) \mHu^H + \vv \vv^H)^{-1} \vv] \|
\leq O(\|\vq(t) - \vqq_k\|), \label{eq:line} \ee as shown in Appendix
A.  Due to \eqref{eq:line} as the number of piecewise linear regions,
i.e., $K$, increases and $\|\vq(t) - \vqq_k\|$ approaches to $0$, the
MSE of the converged adaptive filter more accurately approximates the
MSE of the linear MMSE equalizer.

In the algorithm of Fig. \ref{fig:piece}, the partition of the space
of $\vq \in [0,1]^{N+M-2}$ is fixed, i.e., partitioned regions are
fixed at the start of the equalization, after the VQ algorithm, and we
sequentially learn a different linear equalizer for each region. Since
the equalizers are sequentially learned with a limited amount of data,
these may cause training problems if there is not enough data in each
region. In other words, although one can increase $K$ to increase
approximation power, if there is not enough data to learn the
corresponding linear models in each region, this may deteriorate the
performance. To alleviate this, one can try a piecewise linear model
with smaller $K$ in the start of the learning and gradually increase
$K$ to moderate values if enough data is available.  In the next
section, we examine the context tree weighting method that
intrinsically does such weighting among different models based on
their performance, hence, allowing the boundaries of the partition
regions to be design parameters.
\subsection{Piecewise Linear Turbo Equalization Using Context Trees \label{sec:alg_a}}
\begin{figure}[tb]
\centerline{\epsfxsize=9cm \epsfbox{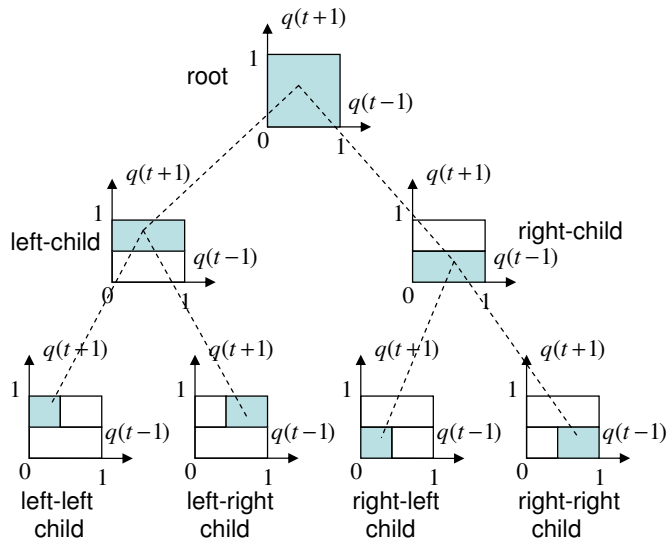}}
\caption{A full binary context tree with depth, $D=2$, with 4 leaves. The leaves of this binary tree partitions $[0,1]^2$, i.e., $[q(t-1)\; \; q(t+1)] \in [0,1]^2$, into 4 disjoint regions. }
\label{fig:tree}
\end{figure}

We first introduce a binary context tree to partition
$[0,1]^{N+M-2}$ into disjoint regions in order to construct a
piecewise linear equalizer that can choose both the regions
as well as the equalizer coefficients in these regions based on the
equalization performance. On a context tree,
starting from the root node, we have a left hand child and a right
hand child.  Each left hand child and right hand child have their own
left hand and right hand children. This splitting yields a
binary tree of depth $D$ with a total of $2^D$ leaves at depth $D$ and
$2^{D+1}-1$ nodes. As an example, the context tree with $D=2$ in
Fig.~\ref{fig:tree} partitions $[0,1]^{2}$, i.e.,
$\vq=[q(t-1),q(t+1)]^T \in [0,1]^{2}$, into $4$
disjoint regions. Each one of these $4$ disjoint regions is assigned
to a leaf on this binary tree. Then, recursively, each internal node
on this tree represents a region (shaded areas in
Fig.~\ref{fig:tree}), which is the union of the regions assigned to
its children.

\begin{figure}[tb]
\centerline{\epsfxsize=13cm \epsfbox{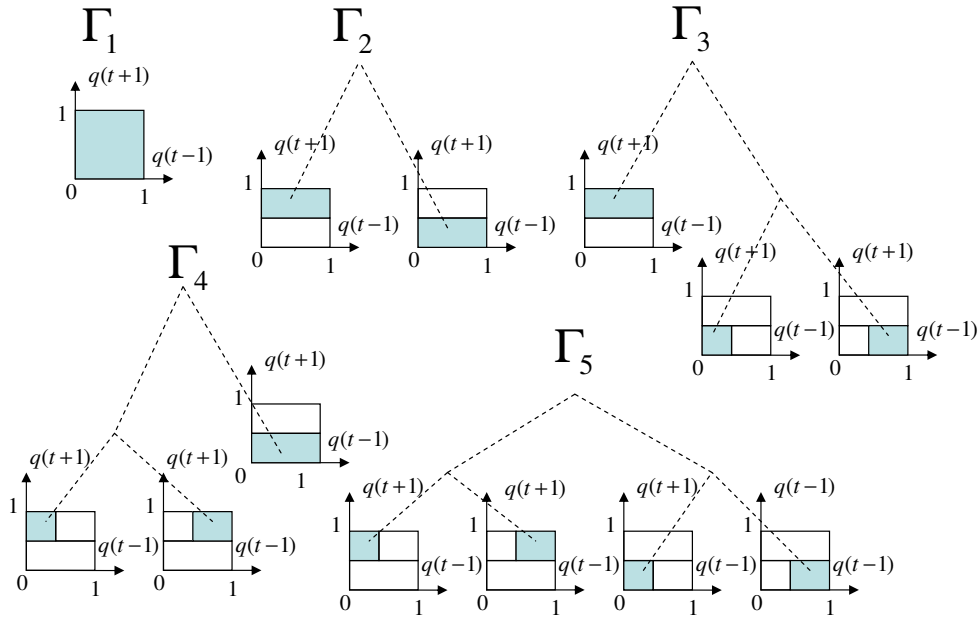}}
\caption{All partitions of $[0,1]^2$ using binary context tree with $D=2$. Given any partition, the union of the regions represented by the leaves of each partition is equal to $[0,1]^2$.}
\label{fig:partition}
\end{figure}

On a binary context tree of depth $D$, one can define a doubly
exponential number, $m\approx 1.5^{2^D}$, of ``complete" subtrees as
in Fig. \ref{fig:partition}. A complete subtree is constructed from a
subset of the nodes of the original tree, starting from the same root
node, and the union of the regions
assigned to its leaves yields $[0,1]^{N+M-2}$. For example,
for a subtree $i$, if the regions assigned to its
leaves are labeled as $\vol_{1,i},\ldots,\vol_{K_i,i}$ where $K_i$ is
the number of leaves of the subtree $i$, then $[0,1]^{N+M-2} =
\bigcup_{k=1}^{K_i}\vol_{k,i}$. Each $\vol_{k,i}$ of the subtree corresponds to a node in the
original tree.

With this definition, a complete subtree with the regions assigned to
its leaves defines a complete ``partition" of
$[0,1]^{N+M-2}$. Continuing with our example, on a binary tree of
depth $D=2$, we have $5$ different partitions of $[0,1]^{2}$ as shown
in Fig. \ref{fig:partition} labeled as $\Pa_1,\ldots,\Pa_5$.
\begin{figure}[h]
{\scriptsize
\begin{tabular}[h]{|l|}
\hline
\textbf{A Pseudo-code of Piecewise Linear Turbo Equalizer Using Context Trees:}\\
\hline
{\bf \% $1$st iteration:} \\ for $t=1,\ldots,n$: \\
\hspace{0.3in}$\hat{x}_{\mathrm{ctw}}(t) = \vw^{(1)T}(t)\vy(t)$, \\
\hspace{0.3in}if $t \leq T$: $e(t) = x(t)-\hat{x}(t)$. \\
\hspace{0.3in}elseif $t > T$: $e(t) = Q(\hat{x}(t))-\hat{x}(t)$, $Q(\cdot)$ is a quantizer. \\
\hspace{0.3in}$\vw^{(1)}(t+1) = \vw^{(1)}(t)+ \mu e(t) \vy(t).$ \\
calculate $\vq(t)$ using the SISO decoder for $t >T$. \\
{\bf \% $m$th iteration:} \\
apply LBG VQ algorithm to $\qts_{t>T}$ to generate $\vqq_k^{(2)}$,
$k=1,\ldots,2^D$ and $\vqq_k^{(2)}$, $k=2^D+1,\ldots,2^{D+1}-1$. \\
if $m == 2$:\\
\hspace{0.3in} for $k=1,\ldots,2^{D+1}-1$: $\vw_k^{(2)}(1) = \vw^{(1)}(n)$, where $\vw^{(1)}(n)$ is from 1st iteration. \\
if $m \geq 3$: \\
\hspace{0.3in}  for $i=1,\ldots,2^{D+1}-1$: $\vw_k^{(m)}(1)= \vw_j^{(m-1)}(n)$, $\vf_k^{(m)}(1)= \vf_j^{(m-1)}(n)$, where $j = \arg\min_i \|\vqq_k^{(m)}-\vqq_i^{(m-1)}\|$. \\
for $k=1,\ldots,2^{D+1}-1$: $A_k(0)=1$, $B_k(0)=1$.   \hfill (line A)  \\

for $t=1,\ldots,T$:  \\
\hspace{0.3in}for $i=1,\ldots,2^{D}$: \\
\hspace{0.6in}$\bar{\vx}(t)\defi [\vec{I}-\mathrm{diag}(\vqq_k^{(m)})]^{1/2} \vx(t)$ \%to consider uncertainty during training\\
\hspace{0.6in}$\vec{l}(1)=i$, where $\vl$ corresponds dark nodes starting from the leaf node $i$\\
\hspace{0.6in}$\eta_1(t)=1/2$. \\
\hspace{0.6in}for $l=2,\ldots,D+1$: \\
\hspace{0.9in}$\eta_l(t) = \frac{1}{2}A_s(t-1) \eta_{l-1}(t)$, where $s$ is a sibling node of $\vec{l}(l)$, i.e., $\vol_s \bigcup \vol_{\vec{l}(l)}= \vol_{\vec{l}(l-1)}$,\\
\hspace{0.9in}$\beta_l(t) = \frac{\eta_l(t) B_{\vec{l}(l)}(t-1)}{A_{\vec{l}(1)}(t-1)}$.   \hfill (line B)\\
\hspace{0.6in}for $l=D+1,\ldots,1$: \\
\hspace{0.9in}$e_{\vec{l}(l)}(t)=x(t)-[\vw^{(m)T}_{\vec{l}(l)}(t)\vy(t)-\vf^{(m)T}_{\vec{l}(l)}(t)\vmx(t)]$, \\
\hspace{0.9in}$B_{\vec{l}(l)}(t) = B_{\vec{l}(l)}(t-1)\exp\left(-c \left\|{e_{\vec{l}(l)}(t)}\right\|^2\right)$,   \% where $c$ is a positive constant,  \hfill (line C)\\
\hspace{0.9in} if $l = D+1$: $A_{\vec{l}(l)}(t)=B_{\vec{l}(l)}(t)$,  \hfill (line D) \\
\hspace{0.9in} else: $A_{\vec{l}(l)}(t) = \frac{1}{2}A_{\vec{l}(l),l}(t-1)A_{\vec{l}(l),r}(t-1)+\frac{1}{2} B_{\vec{l}(l)}(t)$,  \hfill (line E) \\
\hspace{0.9in}\% where $(\vec{l}(l),l)$ and $(\vec{l}(l),r)$ are the left and right hand children of $\vec{l}(l)$, respectively,\\
\hspace{0.9in}$\vw_{\vec{l}(l)}^{(m)}(t+1) = \vw_{\vec{l}(l)}^{(m)}(t)+ \mu e_{\vec{l}(l)}(t) \vy(t)$, $\vf_{\vec{l}(l)}^{(m)}(t+1) = \vf_{\vec{l}(l)}^{(m)}(t)+ \mu e_{\vec{l}(l)}(t) \vmx(t).$ \\

for $t=T+1,\ldots,n$:  \\
\hspace{0.3in}$i = \arg \min_k \|\vq(t)-\vqq_k^{(m)}\|$, $k=1,\ldots,2^D$.\\
\hspace{0.3in}find nodes that $i$ belongs to and store them in $\vec{l}$ starting from the leaf node $i$, i.e., $\vec{l}(1)=i$. \\
\hspace{0.3in}$\eta_1(t)=1/2$, \\
\hspace{0.3in}for $l=2,\ldots,D+1$: \\
\hspace{0.6in}$\eta_l(t) = \frac{1}{2}A_s(t) \eta_{l-1}(t)$.\\
\hspace{0.6in}$\beta_l(t) = \frac{\eta_l(t) B_{\vec{l}(l)}(t-1)}{A_{\vec{l}(1)}(t-1)}$.\\
\hspace{0.3in}$\hat{x}_{\mathrm{ctw}}(t) = \sum_{k=1}^{D+1} \beta_k(t)[\vw^{(m)T}_{\vec{l}(k)}(t)\vy(t)-\vf^{(m)T}_{\vec{l}(k)}(t)\vmx(t)]$, \hfill (line F)\\
\hspace{0.3in}for $l=D+1,\ldots,1$: \\
\hspace{0.6in}$e_{\vec{l}(l)}(t)=Q(\hat{x}_{\mathrm{ctw}}(t))-[\vw^{(m)T}_{\vec{l}(l)}(t)\vy(t)-\vf^{(m)T}_{\vec{l}(l)}(t)\vmx(t)]$, \hfill (line G)\\
\hspace{0.6in}$B_{\vec{l}(l)}(t+1) = B_{\vec{l}(l)}(t-1)\exp\left(-c \left\|{e_{\vec{l}(l)}(t)}\right\|^2\right)$,\\
\hspace{0.6in} if $l = D+1$: $A_{\vec{l}(l)}(t)=B_{\vec{l}(l)}(t)$,\\
\hspace{0.6in} else: $A_{\vec{l}(l)}(t) = \frac{1}{2}A_{\vec{l}(l),l}(t-1)A_{\vec{l}(l),r}(t-1)+\frac{1}{2} B_{\vec{l}(l)}(t)$,\\
\hspace{0.6in}$\vw_{\vec{l}(l)}^{(m)}(t+1) = \vw_{\vec{l}(l)}^{(m)}(t)+ \mu e_{\vec{l}(l)}(t) \vy(t)$, $\vf_{\vec{l}(l)}^{(m)}(t+1) = \vf_{\vec{l}(l)}^{(m)}(t)+ \mu e_{\vec{l}(l)}(t) \vmx(t).$ \\
calculate $\qts_{t >T}$ using the SISO decoder. \\
\hline
\end{tabular}}
\caption{A context tree based turbo equalization. This algorithm requires $O\big(D(M+N)\big)$ computations.
\label{fig:treealg}}
\end{figure}
As in the previous section, we partition regions with the LBG VQ algorithm
and then construct our context tree over
these  regions as follows. Suppose the LBG VQ algorithm is
applied to $\qts$ with $K = 2^D$ to generate $2^D$ regions
\cite{book:gersho}. The LBG VQ algorithm uses a tree notion similar
to the context tree introduced in Fig. \ref{fig:partition} such that
the algorithm starts from a root node which calculates the mean of
all the vectors in $\qts$ as the root codeword, and binary splits
the data as well as the root codeword into two segments.
Then, these newly constructed codewords are iteratively used as
the initial codebook of the split segments. These two codewords
are then split in four and the process is repeated until the
desired number of regions are reached. At the end, this binary
splitting and clustering yield $2^D$ regions with the corresponding
centroids $\vqq_i$, $i=1,\ldots,2^D$, which are assigned to
the leaves of the context tree. Note that since each couple of the
leaves (or nodes) come from a parent node after a binary splitting,
these parent codewords are stored as the internal nodes of the context
tree, i.e., the nodes that are generated by splitting a parent node
are considered as siblings of this parent node where the centroid
before splitting is stored. Hence, in this sense, the LBG VQ algorithm
intrinsically constructs the context tree. However, note that, at each
turn, even though the initial centroids at each splitting directly
come from the parent node in the original LBG VQ algorithm, the final
regions of the leaf nodes while minimizing distortion by iterating
\eqref{eq:vq1} and \eqref{eq:vq2} may invade the regions of the other
parents nodes, i.e., the union of regions assigned to the children of
the split parent node can be different than the region assigned to the
parent node. Note that one can modify the LBG algorithm with the
constraint that the region of the children nodes should be optimized
within the regions of the their parent nodes. However, this constraint may
deteriorate the performance due to more quantization error.

Given such a context tree, one can define $m\approx(1.5)^{2^D}$
different partitions of the space spanned by $\vq$ and construct a
piecewise linear equalizer as in Fig. \ref{fig:piece} for each such
partition. For each such partition, one can train and use a piecewise
linear model.
Note that all these
piecewise linear equalizers are constructed using subsets of nodes
$\rho \in \{ 1,\ldots, 2^{D+1}-1\}$. Hence, suppose we number each
node on this context tree $\rho = 1,\ldots,2^{D+1}-1$ and assign a
linear equalizer to each node as $\hbn =
\vw_{\rho}^T(t)\vy(t)-\vf_{\rho}^T(t)\vmx(t)$. The linear models
$\vw_{\rho}$, $\vf_{\rho}$ that are assigned to node $\rho$, train
only on the data assigned to that node as in Fig. \ref{fig:piece},
i.e., if $\vq(t)$ is in the region that is assigned to the node
$\vol_\rho$ then $\vw_\rho$ and $\vf_\rho$ are updated.  Then, the
piecewise linear equalizer corresponding to the partition ${\Pa}_i = \{\vol_{1,i},\ldots,\vol_{K_i,i}\}$
(where $\bigcup_{l=1}^{K_i}\vol_{k,i} = [0,1]^{N+M-2}$),
say $\hat{x}_{{\Pa}_i}(t)$, is defined such that if $\vq(t) \in
\vol_{k,i}$ and $\rho$ is the node that is assigned to $\vol_{k,i}$
then
\begin{align}
\hat{x}_{{\Pa}_i}(t) & = \hbn \nn \\
                 & = \vw_{\rho}^T(t) \vy(t) - \vf_{\rho}^T(t) \vmx(t). \label{eq:explained}
\end{align}
One of these partitions, with the given piecewise adaptive linear
model $\hat{x}_{{\Pa}_i}(t)$ achieves the minimal loss, e.g., the
minimal accumulated squared error $\sum_{t=1}^n
(x(t)-\hat{x}_{{\Pa}_i}(t))^2$, for some $n$. However, the best piecewise
model with the best partition is not known {\it a priori}.
We next introduce an algorithm
that achieves the performance of the best partition with the best
linear model that achieves the minimal accumulated square-error with
complexity only linear in the depth of the context tree per sample, i.e.,
complexity $O\left(D(2N+M) \right)$ instead of $O\left( (1.5)^{2^D}D
(2N+M) \right)$, where $D$ is the depth of the tree. \\

\noindent
{\bf Remark 2:} We emphasize that the partitioned model that corresponds
to the union of the leaves, i.e., the finest partition, has the finest
partition of the space of variances. Hence, it has the highest number
of regions and parameters to model the nonlinear dependency. However,
note that at each such region, the finest partition needs to train the
corresponding linear equalizer that belongs to that region. As an
example, the piecewise linear equalizer with the finest partition may not
yield satisfactory results in the beginning of the adaptation if there
is not enough data to train all the model parameters. In this sense,
the context tree algorithm adaptively weights coarser and finer models
based on their performance.

To accomplish this, we introduce the algorithm in
Fig. \ref{fig:treealg}, i.e., $\hat{x}_{\mathrm{ctw}}(t)$, that is
constructed using the context tree weighting method introduced in
\cite{willems}. The context tree based equalization algorithm
implicitly constructs all $\hat{x}_{{\Pa}_i}(t)$, $i=1,\ldots,m$,
piecewise linear equalizers and acts as if it had actually
run all these equalizers in parallel on the received data.
At each time $t$, the final estimation $\hat{x}_{\mathrm{ctw}}(t)$
is constructed as a weighted combination
of all the outputs $\hat{x}_{{\Pa}_i}(t)$ of these piecewise linear
equalizers, where the combination weights are calculated proportional
to the performance of each equalizer $\hat{x}_{{\Pa}_i}(t)$ on the
past data. However, as shown in \eqref{eq:explained}, although there
are $m$ different piecewise linear algorithms, at each time $t$, each
$\hat{x}_{{\Pa}_i}(t)$ is equal to one of the $D$ node estimations
to which $\vq(t)$ belongs, e.g., if $\vq(t)$ belongs to the left-left
hand child on Fig. \ref{fig:tree}, $\hat{x}_{{\Pa}_i}(t)$,
$i=1,\ldots,m$, use either the node estimations $\hat{x}_{\rho}(t)$
that belong to the left-left hand child or the left-hand child or the
root. How the context tree algorithm keeps the track of these $m$
piecewise linear models as well as their performance-based combination
weights with computational complexity only linear in the depth of the
context tree is explained in Appendix B.

For the context tree algorithm, since there are no {\it a priori}
probabilities in the first iteration, the first iteration of
Fig. \ref{fig:treealg} is the same as the first iteration of
Fig. \ref{fig:piece}. After the first iteration, to incorporate the
uncertainty during training as in Fig. \ref{fig:piece}, the context
tree algorithm is run by using weighted training data
\cite{kim_singer}. At each time $t > T$, $\hat{x}_{\mathrm{ctw}}(t)$
constructs its nonlinear estimation of $x(t)$ as follows. We first
find the regions to which $\vq(t)$ belongs. Due to
the tree structure, one needs only find the leaf node
in which $\vq(t)$ lies and collect all the parent nodes towards the
root node. The nodes to which $\vq(t)$ belongs  are stored in $\vec{l}$
in Fig. \ref{fig:treealg}. The final estimate
$\hat{x}_{\mathrm{ctw}}(t)$ is constructed as a weighted combination
of the estimates generated in these nodes, i.e., $\hat{x}_\rho(t)$,
$\rho \in \vec{l}$, where the weights are functions of the performance
of the node estimates in previous samples. At each time $t$,
$\hat{x}_{\mathrm{ctw}}(t)$ requires $O(\ln(D))$ calculations to find
the leaf to which $\vq(t)$ belongs. Then, $D+1$ node estimations,
$\hat{x}_\rho(t)$, $\rho \in \vec{l}$, are calculated and the filters
at these nodes should be updated with $O(2N+M)$ computations. The
final weighted combination is produced with $O(D)$
computations. Hence, at each time the context tree algorithm requires
$O(D(2N+M))$ computational complexity. For this algorithm, we have the
following result.  \\ \\
\noindent{\it {\bf Theorem 2:} Let $\xts$, $\nts$ and $\yts$ represent the
transmitted, noise and received signals and $\qts$ represents the
sequence of variances constructed using the {\it a priori}
probabilities for each constellation point produced by the SISO decoder. Let
$\hat{x}_{\rho}(t)$, $\rho = 1,\ldots,2^{D+1}-1$, are estimates of
$x(t)$ produced by the equalizers assigned to each node on the context
tree. The algorithm $\ctr$, when applied to $\yts$, for all $n$
achieves \be \sum_{t=1}^n \big( x(t) - \ctr \big)^2 \leq
\min_{{\Pa}_i} \left\{ \sum_{t=1}^n \big[ x(t)-\pr \big]^2 + 2K_i-1
\right\}, \label{eq:theo1} \ee for all $i$,
$i=1,\ldots,m\approx(1.5)^{2^D}$, assuming perfect feedback in
decision directed mode i.e., $Q(\hat{x}(t))=x(t)$ when $t \geq T$,
where $\hat{x}_{{\Pa}_i}(t)$ is the equalizer constructed as
\[
\hat{x}_{{\Pa}_i}(t) = \hat{x}_{\rho}(t),
\]
$\rho$ is the node  assigned to the volume in ${\Pa}_i =
\{\vol_{1,i},\ldots,\vol_{K_i,i}\}$ such that $\vq(t)$ belongs and
$K_i$ is the number of regions in $\Pa_i$. If the estimation
algorithms assigned to each node are selected as adaptive linear
equalizers such as an RLS update based algorithm, \eqref{eq:theo1} yields
\begin{align}
\sum_{t=1}^n \big[ x(t) - \ctr \big]^2   \leq \min_{i} \Bigg\{ & \min_{\underset{k=1,\ldots,K_i}{\vw_{k,i} \in \mathbbm{C}^N,\vf_{k,i} \in \mathbbm{C}^{N+M-1}}} \sum_{t=1}^n \big[ x(t)-\vw_{s_i(t),i}^T \vy(t)-\vf_{s_i(t),i}^T \vmx(t)]^2 +\nn \\
& O\big( (2N+M) \ln(n) \big) + 2K_i-1 \Bigg\}. \label{eq:theo1_2}
\end{align}
where $s_i(t)$ is an indicator variable for $\Pa_i$ such that if
$\vq(t) \in \vol_{k,i}$, then $s_i(t)=k$.}

An outline of the proof of this theorem is given in Appendix B.  \\ 

\noindent
{\bf
  Remark 3:} We observe from \eqref{eq:theo1} that the context tree
algorithm achieves the performance of the best sequential algorithm
among a doubly exponential number of possible algorithms. Note that
the bound in \eqref{eq:theo1} holds uniformly for all $i$, however the
bound is the largest for the finest partition corresponding to all
leaves. We observe from \eqref{eq:theo1_2} that the context tree
algorithm also achieves the performance of even the best piecewise
linear model, independently optimized in each region, for all $i$ when
the node estimators in each regions are adaptive algorithms that
achieve the minimum least square-error.
\subsubsection{MSE Performance of the Context Tree Equalizer}
To get the MSE performance of the context tree equalizer, we observe
that the result \eqref{eq:theo1_2} in the theorem is uniformly true for
any sequence $\xts$. Hence, as a corollary to the theorem, taking the
expectation of both sides of \eqref{eq:theo1_2} with respect to any distribution on $\xts$ yields the following
corollary: \\
{\it {\bf Corollary:}
\begin{align}
\sum_{t=1}^n E\left\{ \big[ x(t) - \ctr \big]^2\right\}   \leq \min_{{\Pa}_i} \Bigg\{ & \min_{\underset{k=1,\ldots,K_i}{\vw_{k,i} \in \mathbbm{C}^N,\vf_{k,i} \in \mathbbm{C}^{N+M-1}}} \sum_{t=1}^n E\left\{ \big[ x(t)-\vw_{s_i(t),i}^T \vy(t)-\vf_{s_i(t-1),i}^T \vmx(t)]^2\right\}+ \nn \\
& O\big( (N+M) \ln(n) \big) + 2K_i-1 \Bigg\}. \label{eq:cor}
\end{align}}

Note that \eqref{eq:theo1_2} is true for all $i$, and given $i$ for
any $\vw_{k,i}$, $\vf_{k,i}$, $i=1,\ldots,K_i$, i.e.,
\be
\sum_{t=1}^n \big[ x(t) - \ctr \big]^2 \leq \sum_{t=1}^n \big[
  x(t)-\vw_{s_i(t),i}^T \vy(t)-\vf_{s_i(t-1),i}^T \vmx(t) \big]^2 + O\big(
(N+M) \ln(n) \big) + 2K_i-1,\label{eq:corollary}
\ee
since \eqref{eq:theo1_2} is true for the minimizing $i$ and equalizer
vectors.  Taking the expectation of both sides of \eqref{eq:corollary}
and minimizing with respect to $i$ and $\vw_{k,i}$, $\vf_{k,i}$,
$i=1,\ldots,K_i$ yields the corollary.

We emphasize that the minimizer vectors $\vw_{k,i}$ and $\vf_{k,i}$ at
the right hand side of \eqref{eq:cor} minimize the sum of all the
MSEs. Hence, the corollary does not relate the MSE performance of the
CTW equalizer to the MSE performance of the linear MMSE equalizer
given in \eqref{eq:lmmse}.  However, if we assume that the adaptive
filters trained at each node converge to their optimal coefficient
vectors with zero variance and for sufficiently large $D$ and $n$, we
have for piecewise linear models such as for the finest partition \be
\sum_{t=1}^n \big(x(t)- \hat{x}_{{\Pa}_{|K|}}(t)\big)^2 \approx
\sum_{t=1}^n\left\{ \big[ x(t)-\vw_{s_{|K|}(t),|K|,o}^T
  \vy(t)-\vf_{s_{|K}(t),|K|,o}^T \vmx(t) \big]^2
\right\}, \label{eq:approx} \ee where we assumed that, for notational
simplicity, the $|K|$th partition is the finest partition,
$\vw_{s_{|K|}(t),|K|,o}$ and $\vf_{s_{|K|}(t),|K|,o}$ are the MSE
optimal filters (if defined) corresponding to the regions assigned to
the leaves of the context tree. Note that we require $D$ to be large
such that we can assume $\vq(t)$ to be constant in each region such
that these MSE optimal filters are well-defined.  Since
\eqref{eq:theo1_2} is correct for all partitions and for the minimizing
$\vw$, $\vf$ vectors, \eqref{eq:theo1_2} holds for any $\vw$ and
$\vf$'s pairs including $\vw_{s_{|K|}(t),|K|,o}$ and
$\vf_{s_{|K|}(t),|K|,o}$ pair. Since \eqref{eq:theo1_2} in the theorem
is uniformly true, taking the expectation preserves the bound and
using \eqref{eq:approx}, we have \be \frac{1}{n} \sum_{t=1}^n E\left\{
\big[ x(t) - \ctr \big]^2\right\} \leq \frac{1}{n} \sum_{t=1}^n
E\left\{ \big[ x(t)-\vw_{s_{|K|}(t),|K|,o}^T
  \vy(t)-\vf_{s_{|K}(t),|K|,o}^T \vmx(t) \big]^2 \right\} + O\left(
\frac{2^{D+1}}{n}\right), \label{eq:mse_1} \ee
since for the finest partition $K_{|K|}=2^D$.
 Using the MSE
definition for each node in \eqref{eq:mse_1} yields
\begin{align}
& \frac{1}{n} \sum_{t=1}^n  E\left\{ \big[ x(t) - \ctr \big]^2\right\} \nn \\
& \leq \frac{1}{n} \sum_{t=1}^n \left\{ \vw_{s_{|K|}(t),|K|,o}^T \mHu \mQ(t) \mHu^H \vw_{s_{|K|}(t),|K|,o}^*+ \sigma_n^2 \vw_{s_{|K|}(t),|K|,o}^T \vw_{s_{|K|}(t),|K|,o}^* \right\} +  O\left( \frac{2^{D+1}}{n}\right) \label{eq:mse_2}  \\
& \leq  \frac{1}{n} \sum_{t=1}^n \left\{  \min_{\vw,\vf} E\left\{ \big[ x(t)-\vw^T \vy(t)-\vf^T \vmx(t) \big]^2| \vq(t)  \right\} + O\left(\frac{1}{2^D}\right) \right\} + O\left( \frac{2^{D+1}}{n}\right), \label{eq:mse_3}
\end{align}
where \eqref{eq:mse_3} follows from assuming large $D$, the MSE in
each node is bounded as in \eqref{eq:line}. Note that
$O(\|\vq(t)-\tilde{\vq}_k\|)$ at the right hand side of
\eqref{eq:line} can further be upper bounded by $O\left( \frac{1}{2^D}
\right)$ assuming large enough $D$ with the partition given in
Fig.~\ref{fig:tree} since we have $2^D$ regions and $\|\vq(t)\| \leq
\sqrt{N+M-2}$. Hence, as $D\rightarrow \infty$, the context tree
algorithm asymptotically achieves the performance of the linear MMSE
equalizer, i.e., the equalizer that is nonlinear in the variances of
the soft information.
\section{Numerical Examples}
\begin{figure}[tb]
  \centerline{\epsfxsize=8cm \epsfbox{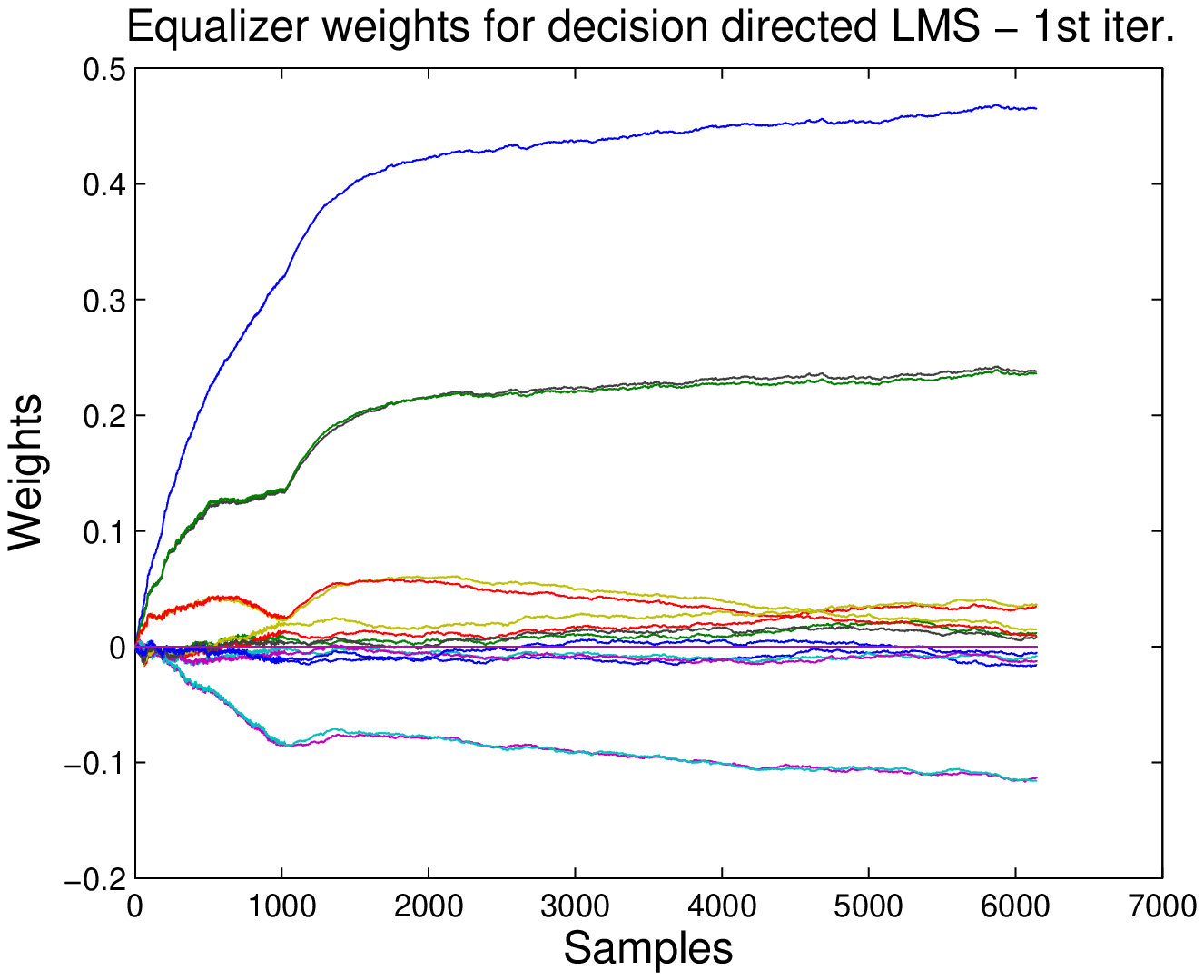}
\hspace{-0.3in} \epsfxsize=8cm \epsfbox{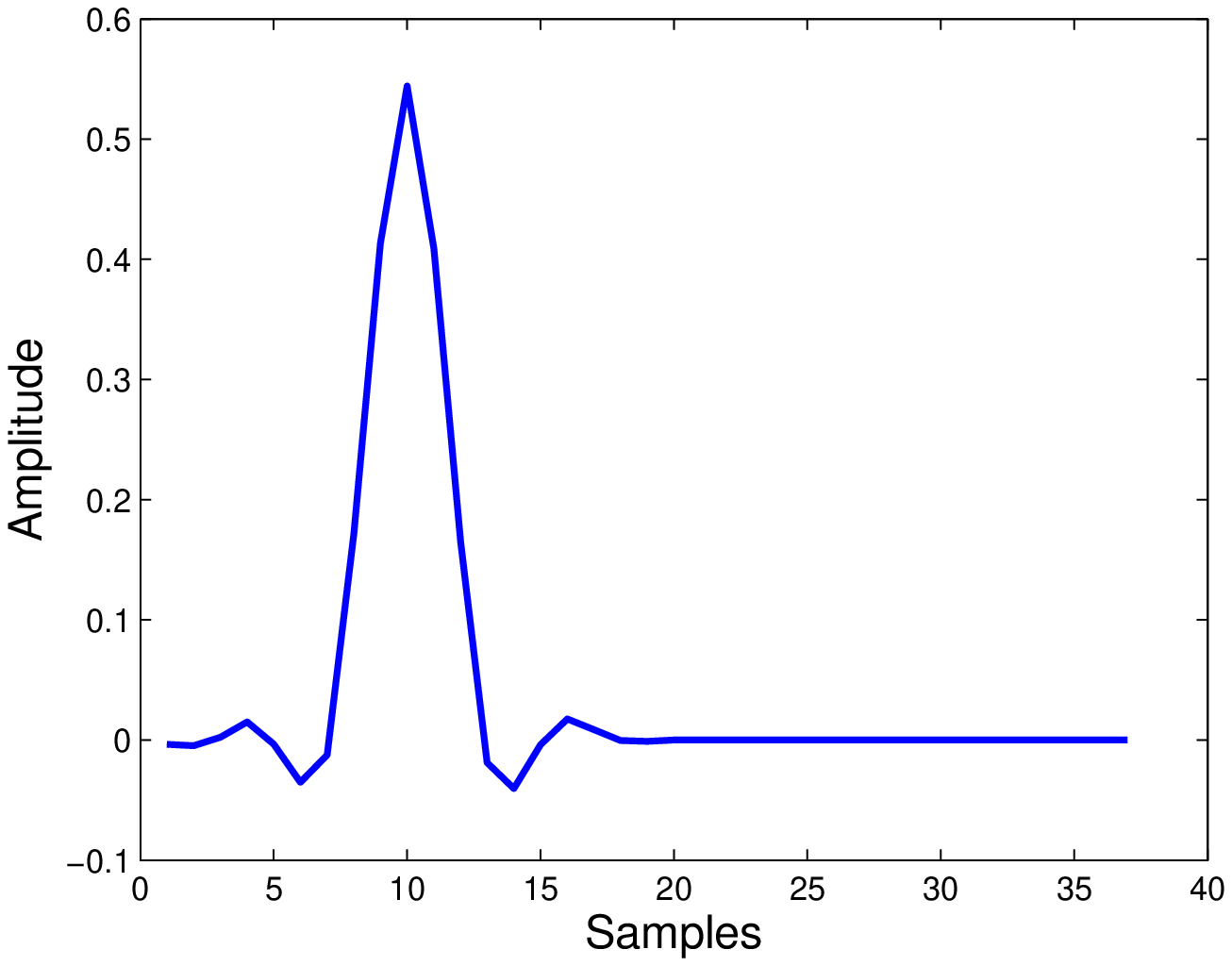}}
\centerline{\hspace{1.6in}(a)\hspace{3.2in}(b)\hfill}
\caption{(a) Ensemble averaged weight vector for the DD LMS algorithm in the first turbo iteration, where $\mu = 0.001$, $T = 1024$ and data length $5120$. (b) Convolution of the trained weight vector of the DD LMS algorithm at sample $5120$ and the channel $\vh$. \label{fig:weight_lms}}
\end{figure}

\begin{figure}[t]
\begin{minipage}[t]{9cm}
\centerline{\epsfxsize=9cm \epsfysize=9cm \leavevmode \epsffile{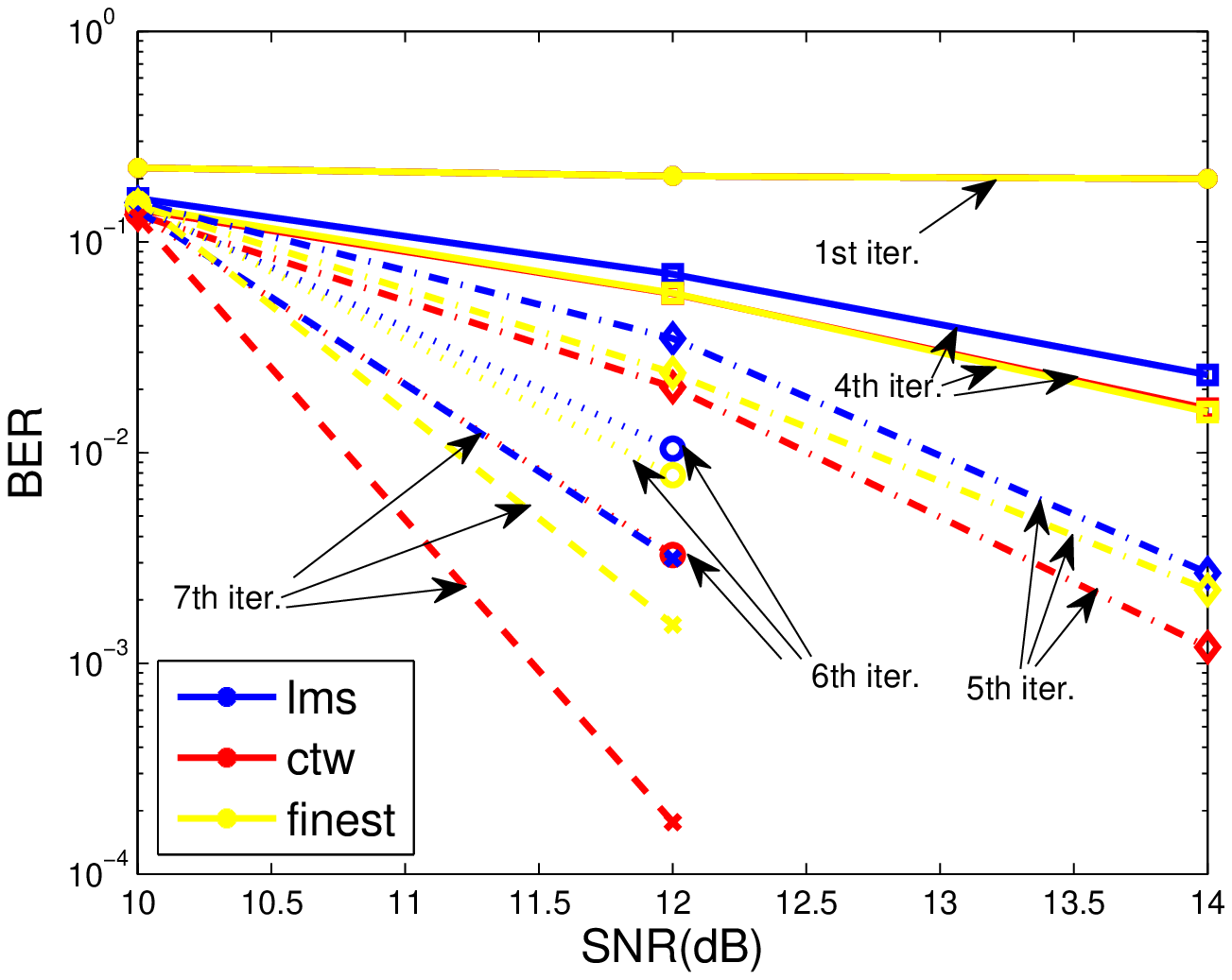}}
\caption{BERs for an ordinary DD LMS algorithm, a CTW equalizer with $D=2$ and tree given in Fig.~\ref{fig:tree}, the piecewise equalizer with the finest partition, i.e., $\hat{x}_{{\Pa}_5}(t)$,  where $\mu=0.001$, $N_1=9$, $N_2=5$, $N+M-1=19$. \label{fig:BER_lti}}
\end{minipage}
\hfill
\hspace{0.08in}
\begin{minipage}[t]{9cm}
\centerline{\epsfxsize=9cm \epsfysize=9cm \leavevmode \epsffile{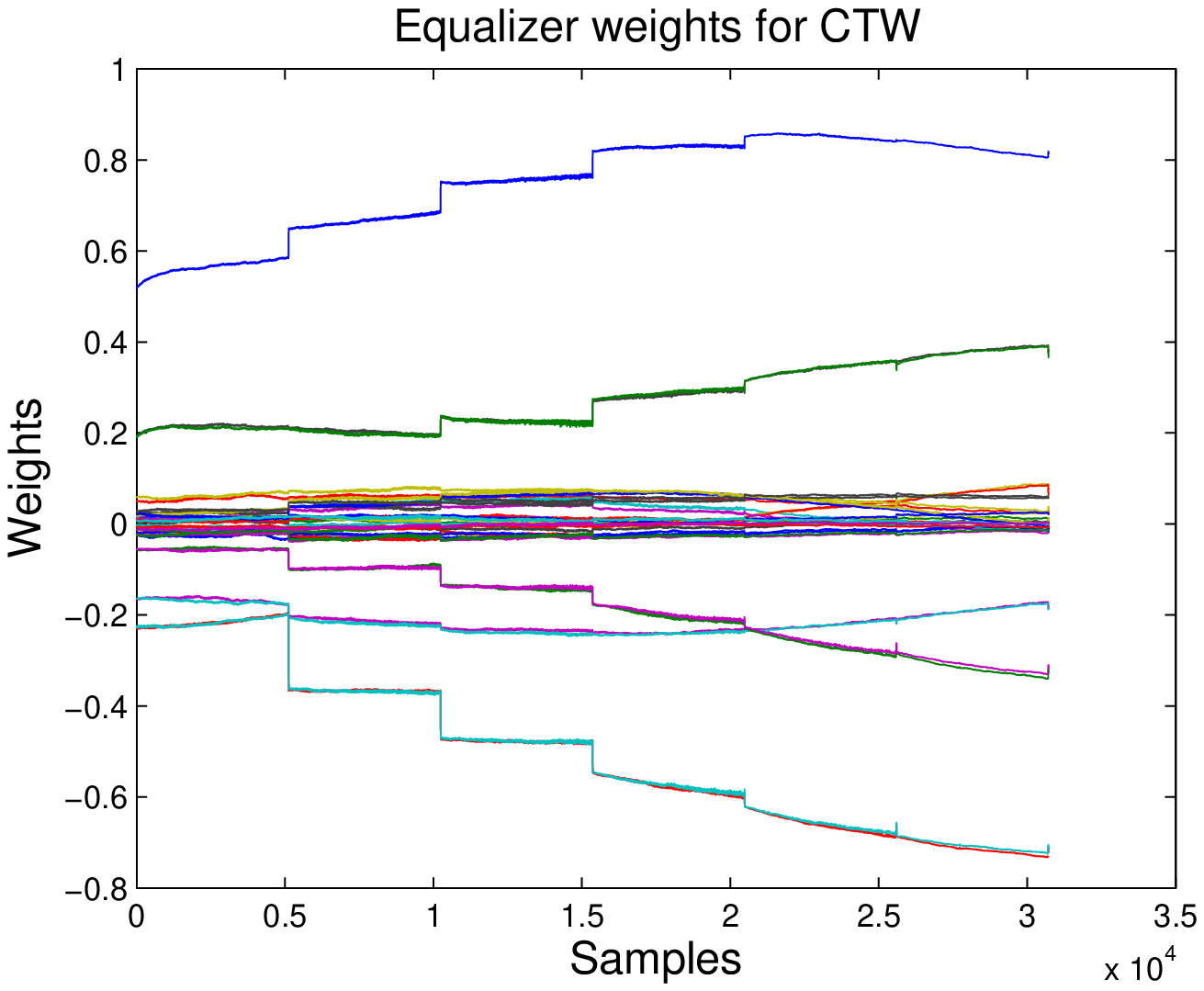}}
\caption{Ensemble averaged combined weight vector for the CTW equalizer over $7$ turbo iterations. Here, we have $\mu = 0.001$, $T = 1024$, data length $5120$ and  $7$ turbo iterations.   \label{fig:weight_ctw}}
\end{minipage}
\hfill
\end{figure}

\begin{figure}[tb]
\centerline{\epsfxsize=8cm \epsfbox{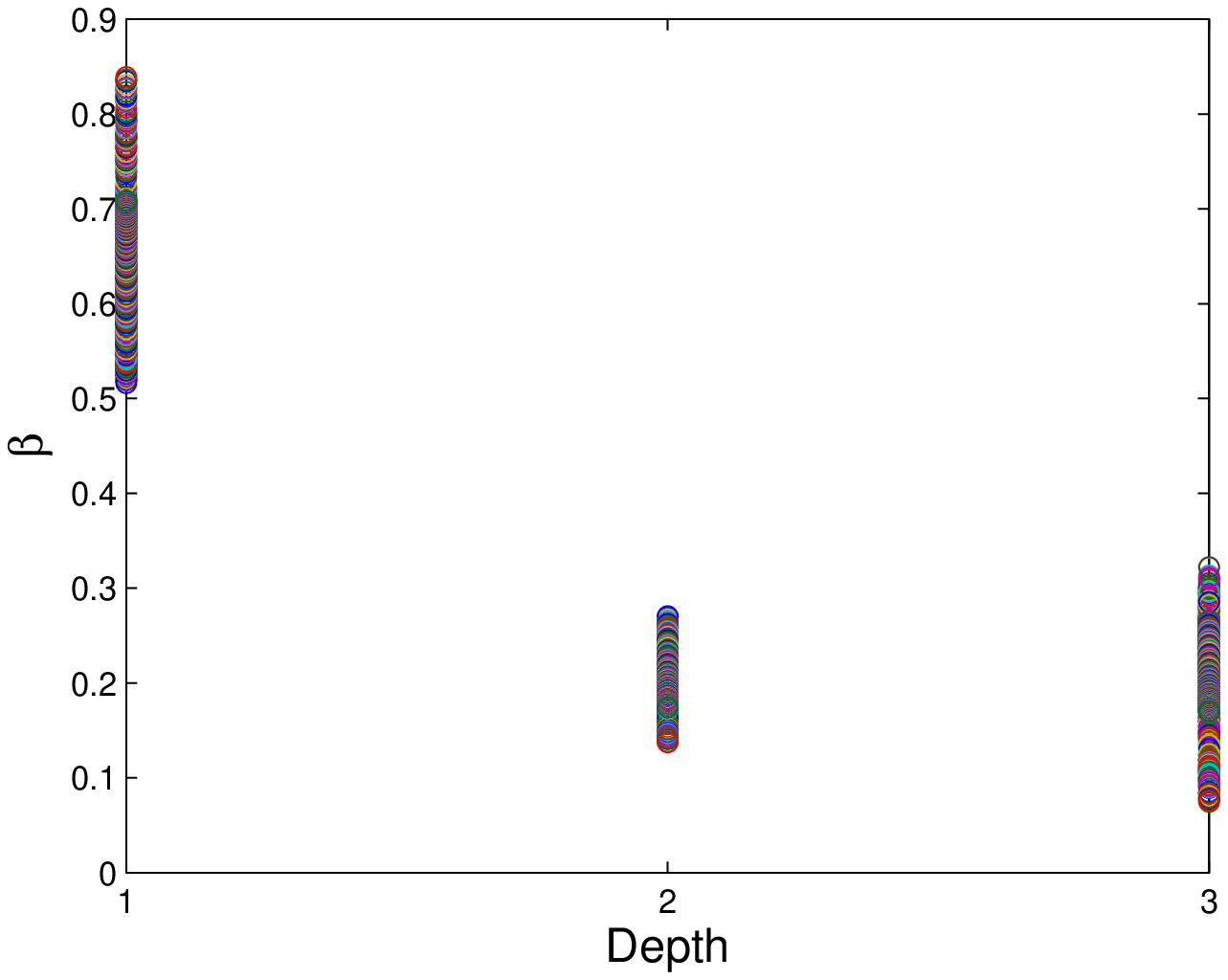}
\hspace{0.3in} \epsfxsize=8cm \epsfbox{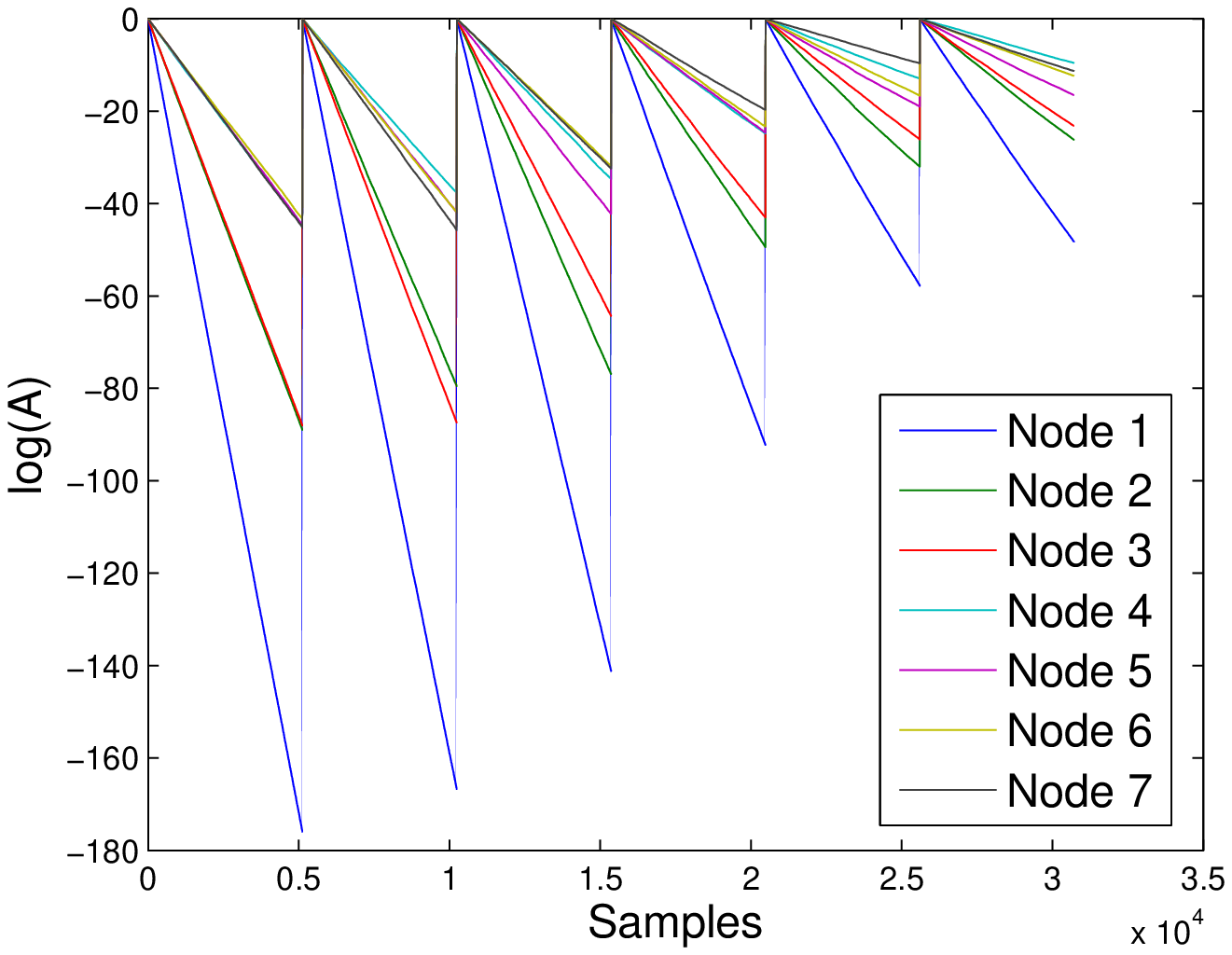}}
\centerline{\hspace{1.6in}(a)\hspace{3.2in}(b)\hfill}
\caption{(a) The distribution of the weights, i.e., values assigned to $\beta_i(t)$, $i=1,2,3$, such that $\beta_i(t)$ belongs to $i$th level.  \label{fig:ctw_learning} (b) Time evaluation of $A_\rho(t)$ which represents the performance of the linear equalizer assigned to node $\rho$. Note that at each iteration, we reset $A_\rho(t)$ since a new tree is constructed using clustering. }
\end{figure}

In the following, we simulate the performance of the algorithms
introduced in this paper under different scenarios. A
rate one half convolutional code with constraint length $3$ and random
interleaving is used. In the first set of experiments, we use the time
invariant channel from \cite{Pr95} (Chapter 10)
\[
\vec{h} = [0.227, 0.46, 0.688, 0.46, 0.227]^T
\]
with the training size $T = 1024$ and data length $5120$ (excluding
training). The BERs and MSE curves are calculated over $20$
independent trials. The decision directed mode is used for all the LMS
algorithms, e.g., for the ordinary LMS turbo equalizer we compete
against and for all the node filters on the context tree. Our
calculation of the extrinsic LLR at the output of the ordinary LMS
algorithm is based on \cite{drost08}. For all LMS filters, we use
$N_1=9$, $N_2=5$, length $N+M-1=19$ feedback filter. The learning
rates for the LMS algorithms are set to $\mu = 0.001$. This learning
rate is selected to guarantee the convergence of the ordinary LMS
filter in the training part. The same learning rate is used directly
on the context tree without tuning. In Fig.~\ref{fig:weight_lms}a, we
demonstrate the time evaluation of the weight vector for the ordinary
LMS turbo equalization algorithm in the first turbo iteration. We also
plot the convolution of the channel $\vh$ and the converged weight
vector of the LMS algorithm at the end of the first iteration in
Fig.~\ref{fig:weight_lms}b.  In Fig.~\ref{fig:BER_lti}, we plot BERs
for an ordinary LMS algorithm, a context-tree equalization algorithm
with $D=2$ given in Fig. \ref{fig:treealg} and the piecewise linear
equalization algorithm with the finest partition, i.e.,
$\hat{x}_{{\Pa}_{|K|}}(t)$, on the same tree. Note that the piecewise
linear equalizer with the finest partition, i.e., $\Pa_5$, in
Fig.~\ref{fig:partition}, has the finest partition with the highest
number of linear models, i.e., $2^D$ independent filters, for
equalization. However, we emphasize that all the linear filters in the
leaves should be sequentially trained for the finest partition.
Hence, as explained in Section~\ref{sec:alg_a}, the piecewise linear
model with the finest partition may yield inferior performance
compared to the CTW algorithm that adaptively weights all the models
based on their performance. We observe that the context tree equalizer
outperforms the ordinary LMS equalizer and the equalizer corresponding
to the finest partition for these simulations. In
Fig.~\ref{fig:weight_ctw}, we plot the weight evaluation of the
context tree algorithm, i.e., the combined weight in line F of
Fig.~\ref{fig:treealg}, to show the convergence of the CTW algorithm.
Note that the combined weight vector for the CTW algorithm is only
defined over the data length period $5120$ at each turbo iteration,
i.e., the combined weight vector is not defined in the training
period. We collect the combined weight vectors for the CTW algorithm
in the data period for all iterations and plot them in
Fig.~\ref{fig:weight_ctw}. This results jumps in the figure, since at
each discontinuity, i.e., after the data period, we switch to the
training period and continue to train the node filters. The context tree
algorithm, unlike the finest partition model, adaptively weights
different partitions in each level. To see this, in
Fig.~\ref{fig:ctw_learning}a, we plot weights assigned to each level
in a depth $D=2$ context tree. We also plot the time evaluation of the
performance measures $A_\rho(t)$ in Fig.~\ref{fig:ctw_learning}b. We
observe that the context tree algorithm, as expected, at the start of
the equalization divides the weights fairly uniformly among the
partitions or node equalizers. However, naturally, as the training
size increases, when there is enough data to train all the node
filters, the context tree algorithm favors models with better
performance.  Note that at each iteration, we reset node probabilities
$A_\rho(t)=1$ since a new tree is constructed using clustering.

\begin{figure}[t]
\begin{minipage}[t]{9cm}
\centerline{\hspace{-0.5in}\epsfxsize=9cm \epsfysize=9cm \leavevmode \epsffile{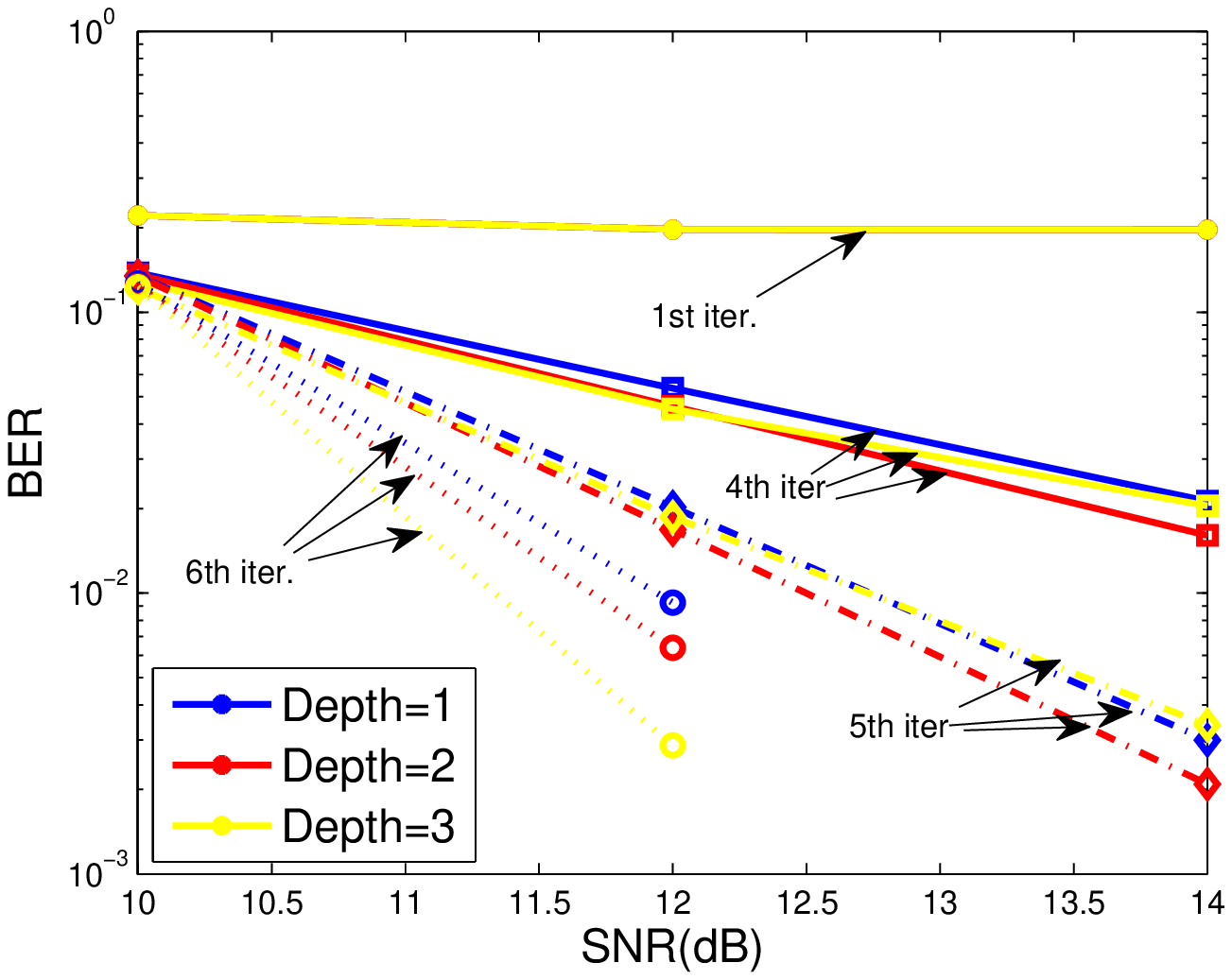}}
\caption{BERs corresponding to CTW equalizers of depth $D=1$, $D=2$ and $D=3$. \label{fig:BER_depth}}
\end{minipage}
\hfill
\begin{minipage}[t]{9cm}
\centerline{\hspace{-0.5in}\epsfxsize=9cm \epsfysize=9cm \leavevmode \epsffile{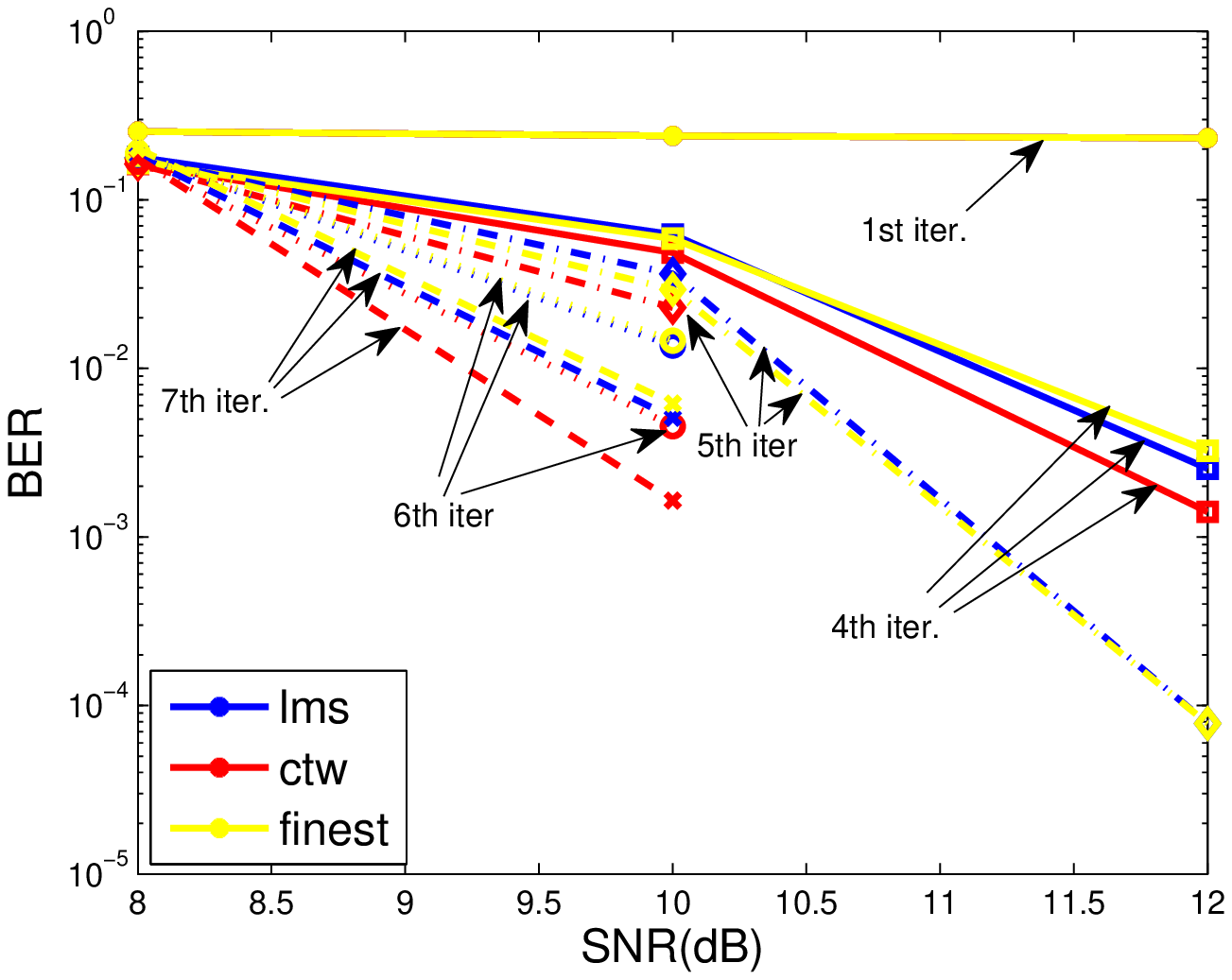}}
\caption{BERs for an ordinary DD LMS algorithm, a CTW equalizer with $D=2$ and tree given in Fig.~\ref{fig:tree}, the piecewise equalizer with the finest partition, i.e., $\hat{x}_{{\Pa}_5}(t)$,  where $\mu=0.001$, $N_1=9$, $N_2=5$, $N+M-1=19$. \label{fig:BER_channel2}}
\end{minipage}
\hfill
\end{figure}

To see the effect of depth on the performance of the context tree
equalizer, we plot the for the same channel, BERs corresponding to
context tree equalizers of depth, $D=1$, $D=2$ and $D=3$ in
Fig.~\ref{fig:BER_depth}. We observe that as the depth of the tree
increases the performance of the tree equalizer improves for these
depth ranges. However, note that the computational complexity of the
CTW equalizer is directly proportional to  depth. As the last set
of experiments, we perform the same set of experiments on a randomly
generated channel of length $7$ and plot the BERs in
Fig.~\ref{fig:BER_channel2}. We observe  similar improvement in BER
for this randomly generated channel for these simulations.
\section{Conclusion}
In this paper, we introduced an adaptive nonlinear turbo equalization
algorithm using context trees to model the nonlinear
dependency of the linear MMSE equalizer on the soft information
generated from the decoder. We use the CTW algorithm to partition the
space of variances, which are time dependent and generated from the
soft information. We demonstrate that the algorithm introduced asymptotically
achieves the performance of the best piecewise linear model defined on this
context tree with a computational complexity only of the order of an
ordinary linear equalizer. We also demonstrate the convergence of the
MSE of the CTW algorithm to the MSE of the linear minimum MSE
estimator as the depth of the context tree and the data length
increase.
\appendix
\noindent {\bf A)} To calculate the difference between the MSE of the
equalizer in \eqref{eq:applms} and the MSE of the linear MMSE
equalizer in \eqref{eq:lmmse}, we start with
\begin{align}
& \| \vw_{k,o}^T \mHu \mQ(t) \mHu^H \vw_{k,o}^*+ \sigma_n^2
\vw_{k,o}^T \vw_{k,o}^* - [1 - \vv^H (\sigma_n^2 \vec{I}+ \mHu \mQ(t)
\mHu^H + \vv \vv^H)^{-1} \vv] \| \nn \\ & = \vw_{k,o}^T \mHu
\Delta\tilde{\mQ}_k \mHu^H \vw_{k,o}^*+ [1 -\vv^H (\sigma_n^2 \vec{I}+
\mHu \tilde{\mQ}_k \mHu^H + \vv \vv^H)^{-1} \vv] - [1 - \vv^H
(\sigma_n^2 \vec{I}+ \mHu \mQ(t) \mHu^H + \vv \vv^H)^{-1} \vv] \nn \\
& = \vw_{k,o}^T \mHu \Delta\tilde{\mQ}_k \mHu^H \vw_{k,o}^* + \vv^H
\left[ (\vec{M}+\mHu \Delta\tilde{\mQ}_k \mHu^H)^{-1}-\vec{M}^{-1}
\right] \vv ,\label{eq:diff}
\end{align}
where $\Delta\tilde{\mQ}_k \defi \mQ(t) -\tilde{\mQ}_k$ (and the time
index in $\Delta \tilde{\mQ}_k$ is omitted for presentation purposes)
and $\vec{M} \defi (\sigma_n^2 \vec{I}+ \mHu \tilde{\mQ}_{k} \mHu^H +
\vv \vv^H)$. To simplify the second term in \eqref{eq:diff}, we use
the first order expansion from the Lemma in the last part of 
Appendix A to yield
\begin{align}
& \vv^H(\vec{M}+\mHu \Delta\tilde{\mQ}_k \mHu^H)^{-1}\vv \\
& = \vv^H \vec{M}^{-1}\vv+
\mathrm{tr} \left\{ \nabla_{\Delta\tilde{\mQ}_k}^H \left[\vv^H (\vec{M}+\mHu \Delta\tilde{\mQ}_k
\mHu^H)^{-1}\vv \right]|_{\Delta\tilde{\mQ}_k = \vec{0}} \Delta\tilde{\mQ}_k \right\}+O(\mathrm{tr}[ \Delta\tilde{\mQ}_k^H \Delta\tilde{\mQ}_k]), \nn \\
& = \vv^H \vec{M}^{-1}\vv +\mathrm{tr}( \vec{M}^{-1}\vv \vv^H \vec{M}^{-1} \Delta\tilde{\mQ}_k)+ O(\mathrm{tr}[ \Delta\tilde{\mQ}_k^H \Delta\tilde{\mQ}_k]), \label{eq:taylor}
\end{align}
around $\Delta\tilde{\mQ}_k = \vec{0}$. Hence using \eqref{eq:taylor}
in \eqref{eq:diff} yields
\begin{align*}
& \vw_{k,o}^T \mHu \Delta\tilde{\mQ}_k \mHu^H \vw_{k,o}^* + \vv^H \left[ (\vec{M}+\mHu \Delta\tilde{\mQ}_k \mHu^H)^{-1}-\vec{M}^{-1} \right] \vv \\
& = \vw_{k,o}^T \mHu \Delta\tilde{\mQ}_k \mHu^H \vw_{k,o}^* + \mathrm{tr}( \vec{M}^{-1}\vv \vv^H \vec{M}^{-1} \Delta\tilde{\mQ}_k) + O(\|\vq(t)-\vqq_k\|^2) \leq O(\|\vq(t)-\vqq_k\|),
\end{align*}
where the last line follows from the Schwartz inequality. $\Box$\\
\noindent 
{\bf Lemma:} We have 
\be
\nabla_{\Delta\tilde{\mQ}_k} \vv^H (\vec{M}+\mHu \Delta\tilde{\mQ}_k
\mHu^H)^{-1}\vv = (\vec{M}+\mHu \Delta\tilde{\mQ}_k \mHu^H)^{-1} \vv \vv^H (\vec{M}+\mHu \Delta\tilde{\mQ}_k \mHu^H)^{-1}. \label{eq:grad}
\ee
{\it Proof:} To get the gradient of $(\vec{M}+\mHu
\Delta\tilde{\mQ}_k \mHu^H)^{-1}$ with respect to
$\Delta\tilde{\mQ}_k$, we differentiate the identity $(\vec{M}+\mHu
\Delta\tilde{\mQ}_k \mHu^H)^{-1} (\vec{M}+\mHu\Delta\tilde{\mQ}_k \mHu^H)=\vec{I}$ with respect to $(\Delta\tilde{\mQ}_k)_{a,b}$, i.e., the $a$th and $b$th element of the matrix $\Delta\tilde{\mQ}_k$ and obtain
\[
\frac{\partial (\vec{M}+\mHu \Delta\tilde{\mQ}_k \mHu^H)^{-1}}{\partial (\Delta\tilde{\mQ}_k)_{a,b}} (\vec{M}+\mHu \Delta\tilde{\mQ}_k \mHu^H)+(\vec{M}+\mHu \Delta\tilde{\mQ}_k \mHu^H)^{-1} (\mHu \vec{e}_a \vec{e}_b^T \mHu^H) = \vec{0},
\]
where $\vec{e}_a$ is a vector of all zeros except a single 1 at $a$th entry. This yields
\begin{align}
\frac{\partial \vv^H (\vec{M}+\mHu \Delta\tilde{\mQ}_k \mHu^H)^{-1}\vv}{\partial (\Delta\tilde{\mQ}_k)_{a,b}} & = \vv^H (\vec{M}+\mHu \Delta\tilde{\mQ}_k \mHu^H)^{-1} \mHu \vec{e}_a \vec{e}_b^T \mHu^H(\vec{M}+\mHu \Delta\tilde{\mQ}_k \mHu^H)^{-1}\vv \nn  \\
& = \mathrm{tr}\left\{ \vec{e}_b \mHu^H(\vec{M}+\mHu \Delta\tilde{\mQ}_k \mHu^H)^{-1}\vv \vv^H (\vec{M}+\mHu \Delta\tilde{\mQ}_k \mHu^H)^{-1} \mHu \vec{e}_a  \right\}, \label{eq:lemma}
\end{align}
which yields the result in \eqref{eq:grad} since \eqref{eq:lemma} is the
$(b,a)$th element of the matrix in \eqref{eq:grad}.$\Box$ \\

\noindent
{\bf B) Outline of the proof of the theorem 2:} The proof of the theorem
follows the proof of the Theorem 2 of \cite{Ko07} and Theorem 1 of
\cite{sinfed}.  Hence, we mainly focus on differences.

Suppose we construct $\hat{x}_{{\Pa}_i}(t)$, $i=1,\ldots,m$ and
compute weights \be c_i(t) \defi \frac{2^{-C(\Pa_i)} \exp\{-a
  \sum_{r=1}^{t-1} [x(r)-\hat{x}_{{\Pa}_i}(r)]^2\}}{\sum_{j=1}^m
  2^{-C(\Pa_j)} \exp\{-a \sum_{r=1}^{t-1}
  [x(r)-\hat{x}_{{\Pa}_j}(r)]^2\}}, \label{eq:weights} \ee where $0<
C(\Pa_j) \leq 2K_j-1$ are certain constants that are used only for proof
purposes such that $\sum_{k=1}^m 2^{-C(\Pa_j)} = 1$ \cite{willems} and
$a$ is a positive constant set to $a = \frac{1}{2} = \frac{1}{2 |\max \xts|^2}$ \cite{sinfed}. If we define $\hat{x}(t)
\defi \sum_{k=1}^m c_k(t) \hat{x}_{{\Pa}_k}(t)$, then it follows from
Theorem 1 of \cite{sinfed} that
\[
\sum_{t=1}^n [x(t)-\hat{x}(t)]^2 \leq \sum_{t=1}^n [x(t)-\hat{x}_{{\Pa}_i}(t)]^2 + O(K_i)
\]
for all $i=1,\ldots,m$. Hence, $\hat{x}(t)$ is the desired
$\ctr$. However, note that $\hat{x}(t)$ requires output of $m$
algorithms and computes $m$ performance based weights in
\eqref{eq:weights}. However, in $\hat{x}(t)$ there are only $D$
distinct node predictions $\hat{x}_\rho(t)$ that $\vq(t)$ belongs to
such that all the weights with the same node predictions can be merged
to construct the performance weighting.  It is shown in \cite{Ko07}
that if one defines certain functions of performance for each node as
$A_\rho(t)$, $B_\rho(t)$ that are initialized in (line A) and updated
in (line C), (line D), (line E) of Fig. \ref{fig:treealg}, then the
corresponding $\hat{x}(t)$ can be written as
$\hat{x}(t)=\sum_{l=1}^{D+1} \beta_k(t) \hat{x}_{\vec{l}(l)}(t)$,
where $\vec{l}$ contains the nodes that $\vq(t)$ belongs to and
$\beta_k(t)$ are calculated as shown in (line B) of
Fig. \ref{fig:treealg}. Hence, the desired equalizer is given by $\ctr
= \sum_{l=1}^{D+1} \beta_k(t) \hat{x}_{\vec{l}(l)}(t)$, which requires
computing $D+1$ node estimations and updates only $D+1$ node
equalizers at each time $t$ and store $2^{D+1}-1$ node weights. This
completes the outline of the proof of
\eqref{eq:theo1}. To get the corresponding result in
\eqref{eq:theo1_2}, we define the node predictors as the LS
predictors such that
\be
\MB{cc} \vw_{\rho}(t) & \vf_{\rho}(t) \ME^T  = \vec{M}^{-1}(t-1)\vec{p}(t-1), \mbox{   } \vec{M}(t-1) \defi \left( \sum_{r=1}^{t} \vec{d}(r-1) \vec{d}(r-1)^Ts_{\rho}(r)  + \delta \vec{I} \right) \label{eq:lin}
\ee
and $ \vec{p}(t-1) \defi \sum_{r=1}^{t-1} Q(\hat{y}(r))
\vec{d}(r-1)s_{\rho}(r)$, where $\vec{d}(r) \defi \MB{cc}\vec{y}(r) &
\vmx(r) \ME^T$, $s_{\rho}(r)$ is the indicator variable for node
$\rho$, i.e., $s_{\rho}(r)=1$ if $\vq(r) \in
\vol_{\rho}$ otherwise $s_\rho(r)=0$. The
affine predictor in \eqref{eq:lin} is a least squares predictor that
trains only on the observed data $\yts$ and $\{ \bar{x}(t)\}$ that
belongs to that node, i.e., that falls into the region
$\vol_{\rho}$. Note that the update in
\eqref{eq:lin} can be implemented with $O(1)$ computations using
fast inversion methods \cite{fastRLS}. The RLS algorithm is shown to achieve the excess loss given in \eqref{eq:theo1_2} as shown in \cite{merhav}. $\Box$

\bibliographystyle{IEEEbib}
\bibliography{bibfile2}
\end{document}